\documentclass[12pt]{article}
\usepackage{epsfig}
\usepackage{amsfonts}
\usepackage{amsmath}
\usepackage{amssymb}
\usepackage{float}
\begin{document}
\title{
Analyzing the contribution of individual resonance poles of the
$S$-matrix to the two-channel scattering
}
\author{S. A. Rakityansky$^{*}$ and N. Elander$^{**}$\\
$^{*}${\small\it Dept. of Physics, University of Pretoria, 
Lynnwood Road, Pretoria 0002, South Africa}\\
$^{**}${\small\it Div.of Molecular Physics, Dept. of Physics, Stockholm 
University, Stockholm, SE-106 91, Sweden}
}
\maketitle
\begin{abstract}
\noindent
A two-channel problem is considered within a method based on first
order differential equations that are equivalent to the corresponding
Schr\"odinger equation but are more convenient for dealing with
resonant phenomena. Using these equations, it is possible to directly
calculate the Jost matrix for practically any complex value of the
energy. The spectral points (bound and resonant states) can therefore
be located in a rigorous way, namely, as zeros of the Jost matrix
determinant.  When calculating the Jost matrix, the differential
equations are solved and thus, at the same time, the wave function is
obtained with the correct asymptotic behavior that is embedded in the
solution analytically.  The method offers very accurate way of
calculating not only total widths of resonances but their partial
widths as well. For each pole of the $S$-matrix, its residue can be
calculated rather accurately, which makes it possible to obtain the
Mittag-Leffler type expansion of the $S$-matrix as a sum of the
singular terms (representing the resonances) and the background term
(contour integral).  As an example, the two-channel model by Noro and
Taylor is considered. It is demonstrated how the contributions of
individual resonance poles to the scattering cross section can be
analyzed using the Mittag-Leffler expansion and the Argand plot
technique. This example shows that even poles situated far away from
the physical real axis may give significant contributions to the cross
section.
\end{abstract}
\vspace{.5cm}
\noindent
PACS number(s): {03.65.Nk, 03.65.-w, 31.15.-p, 34.50.-s}\\[5mm]
Published in:\quad
{\large\bf Int.J.Quant.Chem.,106(2006)1105-1129}

\section{Introduction}

Mankind has always been fascinated by striking phenomena such as
thunderstorms, earthquakes, and comets.  As a part of mankind,
scientists also are more attracted to drastically rather than smoothly
varying quantities.  Quantum physicists considering collisions of the
molecular, atomic or sub-atomic particles, when find irregularities in
their energy dependencies, try to assign certain quantum numbers to
the outstanding features.  These and other phenomena can often be
regarded as resonances.  Friedmann and Truhlar point out that
``Probably the most striking phenomenon in the whole range of
scattering experiments is the resonance'' \cite{R:FriedmanTRu91}.
However, not all irregularities in the scattering data are true
resonances \cite{R:AlonsZare2002}.

Early studies of the energy dependence of nuclear scattering that led
Bohr \cite{R:NBohrNatu37} to the idea of the compound nucleus (for
more elaborate discussions see, for example, Ref. \cite{R:Bilpuch66}),
inspired similar ideas to appear in chemical physics.  The notion of
collision complex, used in atomic and molecular physics as well as
chemistry, was developed in theoretical studies of the early 1970s
\cite{R:LevinWu71,R:SchartzKuper73} (see also Ref.
\cite{R:Schatz2000}).

However, it took about 20 years for indirect experimental evidence of
the existence of such complexes to appear
\cite{R:NeumarkWo85,R:IMwaller90}.  The first direct observation of an
isolated resonance in chemical reactions only appeared around 2000
\cite{R:Schatz2000,R:SkodjeSkouJCP2000}. Still it is not absolutely
clear that the results of Ref. \cite{R:SkodjeSkouJCP2000} concerning
the existence of the $FHD$ collision complex in the $F + HD \to FHD
\to FH + D$ reaction, give an undoubted evidence. For example, in
Ref.\cite{R:FardaySoc04} the doubts are expressed in terms of the
definition of a resonance, which is rigorously defined as a quantum
state that at large distances is a purely outgoing
wave. Mathematically, this implies that the matrix element of the
operator that transforms the incoming wave into the outgoing wave (the
$S$-operator), has a pole at the resonance (complex valued) energy
\cite{R:Newtonbook,R:Taylorbook}.

A theoretical analysis of intermediate states, in the collisions like
$F+HD$, will eventually need more than one electronic potential
energy surface, a feature which is not easily
achieved\cite{R:MillardAManol00}.  It is thus, from a theoretical
point of view, reasonable to begin with atom ion collisions, which can
be represented by a set of coupled potential energy curves.

Looking back in the literature, we find that a few resonances were
found in a theoretical study of the $N^{3+} + H \to N^{2+} + H^{+}$
reaction by McCarroll and Valrion\cite{R:McCarrollVa80}. B\'ar\'any
{\sl et al.} \cite{R:BaranyBRrand83} and Rittby {\sl et al.}
\cite{R:RittbyEla84} extended this work and could assign complex
eigenenergies and their rovibrational quantum numbers to not only the
resonances found in \cite{R:McCarrollVa80}. These resonances were, as
in the work by Elander and coworkers
\cite{R:RittbyElBr89,R:Krylninni89,R:AlferovaNe02,R:AlferovaNe04},
forming a sting that started with a narrow resonance and propagated out
in the complex energy plane.

The group lead by Shimakura, studied several charge-transfer reactions
with the initial channel $N^{5+} + H$ (see Refs.
\cite{R:ShimakuraKim91,R:KitaGo98,R:SuzuliShim98,R:SuzukiShi01}) while
Bacchys-Montabonel and Ceyzeriat \cite{R:BacchusCey98} did
calculations for the ion-atom collision $Si^{4+} + He$.  Both groups
predicted resonances in these reactions. Experimentalists at Oak Ridge
made an effort to find evidence for these type of resonances
\cite{R:HavenerHuq89,R:HuqHav89} but the obtainable energy range and
resolution did not permit their detection. The conclusion was that due
to extremely high Coulomb barrier the relative energy ($\sim$meV) at
which these resonances appear, could never be achieved in collision
experiments.

The interest to the resonant charge-transfers and other similar
processes did not wane. The new idea is to observe the collisions
between particles moving in the same beam, so that their relative
kinetic energy is very low. Electrostatic storage rings for this
purpose are currently being built. For example, the Desiree project in
Stockholm \cite{R:DESIREE} is planned to have the possibility of merging
a stored multiply charged ion beam with a beam of neutral atoms.
Similar experiments are also planned at other facilities.

In the theoretical studies, the most ubiquitous formula associated
with resonance phenomena, is the one bearing the names of Breit and
Wigner. It was invented for phenomenological description of the peaks
in the neutron scattering cross section \cite{R:BreitWigner36} and is
valid in the neighborhood of an isolated resonance.  From textbooks, we
learn that if a resonance is embedded in a direct scattering
continuum, the cross section profile does not have a Lorentian shape
as predicted by the Breit-Wigner formula, but should rather be
described by the more general Butler-Fano parameters \cite{R:Ufano61}.

The original ideas of Breit and Wigner \cite{R:BreitWigner36} and Fano
\cite{R:Ufano61} may make us believe that resonances only give a
signature to a cross section as peaks or similar
irregularities. However, model potential studies of the
Schr{\"o}dinger spectrum \cite{R:Rittby82,R:Abramov01} found that the
$S$-matrix poles associated with resonances, form long (most likely
infinite) strings in the complex energy plane. The influence of these
poles on the cross section could be more complicated than simple
generating of the peaks or bumps. The resonances may interfere with
each other and thus cause an intricate energy dependence of the
cross section.

The most controversial example of such intricate interference of
resonance states is the problem of barrier states in chemical physics.
According to Friedmann and Truhlar, ``metastable states associated
with barriers, are associated with poles of the S-matrix just as
definitely as are trapped states associated with standing waves in a
well'' and these metastable states are the bottle necks for chemical
reactions (see Refs. \cite{R:FriedmanTRu91} and \cite{R:TruhlarGa96}).

In relation to barrier scattering reactions like $H+HD\to D+H_2$,
these ideas are discussed by Althrope {\sl et al.}
\cite{R:AlthorpeaNat2002}, where they report ``an intriguing
forward-scattering peak'' observed at a collision energy 
$\sim$1.64\,eV in a state-to-state differential cross section. From the
point of view of Friedmann and Truhlar \cite{R:FriedmanTRu91}, this
observation could be a manifestation of quantum barrier states
suggested by them. This is opposed by Manolopoulos
\cite{R:ManolopoulosNa2002} who claims that isolated (i.e. true)
resonances and barrier type resonances are ``mathematically different
and the physical implications of this mathematical difference are too
significant to regard the two situations as a manifestation of the
same phenomenon'' \cite{R:AlthorpeaNat2002}.

Other similar results on ``quantized bottleneck state(s)'' are
reported in Ref. \cite{R:HarichDaieaNa2002}.  It should also be
mentioned that the quantized bottleneck ideas do have implications on
the understanding of the so called transition states discussed in
chemical physics ( See Ref. \cite{R:LaiderKing83} for a review).

The controversies, like the barrier states, can only be resolved if we
reach a clear understanding of how much a given isolated or barrier
type resonance influences the cross section. The importance of this is
emphasized by the authors of Refs.
\cite{R:FriedmanTRu91,R:TruhlarGa96,R:AlthorpeaNat2002,R:ManolopoulosNa2002,R:HarichDaieaNa2002}
and many others. A number of methods were developed for the purpose of
analyzing the roles of individual resonances in the scattering cross
section. A comparison between the Mittag-Leffler expansion approach
(see, for example, Ref.\cite{R:RittbyElBr89}), the Siegert
pseudo-state method \cite{R:TON98}, and the complex Kohn variational
method was recently published in Ref. \cite{R:ElaAferOstro04}.
Based on the Titchmarsh-Weyl theory, the authors of Ref.\cite{Brandas}
developed a method for decomposing the scattering information into the 
resonance and background contributions.
One of us (N.E.)  has earlier analyzed the Mittag-Leffler method to
understand its computational properties
\cite{R:Krylninni89,R:AlferovaNe02,R:AlferovaNe04}.  In
Ref. \cite{R:AlferovaNe04}, we also investigated the influence of a
string of resonance poles yielding somewhat surprising results.  

One drawback of the Mittag-Leffler expansion method is that it may be
hard to apply it to the systems involving more than two particles in
at least one of the channels.  In this respect the complex variational
Kohn method, as described by Nuttal and Cohen \cite{R:NuttalCohen69},
Rescigno and McCurdy\cite{R:ResciMcCu85}, and recently in the review
by Moiseyev \cite{MoiRev98}, has an advantage. It can be formulated in
terms of matrix elements of the interaction potential and the
complex eigenvalues of the corresponding analytically continued
Hamiltonian \cite{MoiRev98}. In this way the Kohn method is applicable
whenever a basis set representation of the problem is available.  It can
therefore be extended to the 3-D problems such as, for example, the
collisions complex $FHD$ etc.
\cite{R:NeumarkWo85,R:IMwaller90,R:SkodjeSkouJCP2000}.  However, being
based on the basis expansion, the complex variational Kohn method has
its disadvantages as well. These  are related to the problem of
convergence and completeness of the expansion. It is thus of
importance to compare and analyze the three above mentioned methods in
parallel to understand and utilize their respective powers in
analyzing and predicting experimental results.

As a prerequisite for the analysis of the cross sections of chemical
reactions, one needs a reliable method for solving the multi-channel
Schr\"odinger problem. As we already mentioned, in theoretical
analysis of atomic complexes, like $F + HD$, it is very common to
represent the atom-ion interaction by a set of coupled potential
energy curves.

In this paper,
we adopt a mathematically rigorous definition of a resonance, namely,
as a point on the unphysical sheet of complex energy surface at which
the Jost-matrix determinant is zero. At such points, the $S$-matrix
has poles since this determinant is in the denominators of all its
matrix elements. The so-called redundant (non-resonance) poles that
the $S$-matrix may also have (see, for example, Ref.\cite{baz}),
are thus avoided since they are associated with singularities of the
numerators of its matrix elements. This mathematically-inclined
definition is consistent with physics in that at thus defined
resonance points the wave function has the Gamow-Siegert asymptotics
(pure outgoing waves), but it does not say anything about observable
irregularities of the cross section. Therefore the term ``resonance''
has here a more general meaning not limited to long-lived (narrow)
states. 

In the paper, we describe a mathematically rigorous and numerically
accurate and stable method for solving the two-channel (generally,
$N$-channel) quantum mechanical problem. For the last ten years
various aspects of this method were developed in Refs.
\cite{rakpup,nuovocim,exactmethod,nnn-nnnn,partialwaves,singular,cplch,Massen,Masui,Masui1,Masui2,nano1,nano2}.
Here we suggest an approach to analyzing the contribution of resonance
$S$-matrix poles to the scattering picture.  Our approach is based on
a Mittag-Leffler type expansion of the $S$-matrix, for which we
calculate the residues of the $S$-matrix at the resonance poles and
its Cauchy-type contour integral. The roles played by individual
resonances are deduced using the Argand plot technique. Practical
application of the method is demonstrated by the example of the Noro
and Taylor two-channel model \cite{NoroTaylor}.

\section{Two-channel Schr\"odinger equation}
Let us consider a quantum mechanical two-body problem which, after
separation of the motion of its center of mass, is reduced to an
effective problem of one body whose dynamics is governed by the
Hamiltonian
\begin{equation}
\label{Hamiltonian}
   H=H_0+U+h\ ,
\end{equation}
where the terms on the right hand side describe its free motion
($H_0$), the interaction forces ($U$), and the internal dynamics in
the body ($h$), respectively. Since our main objective is to study how
the resonance poles of the $S$-matrix manifest themselves in
experimentally observable cross sections, we avoid unnecessary
complications associated with spins of the particles and with
long-range Coulomb forces. Furthermore, for the sake of clarity, we
consider here a two-channel system, although all the equations can be
easily generalized for an arbitrary number of channels (see
Ref. \cite{cplch}) as well as for charged particles with non-zero
spins (see Refs.  \cite{exactmethod,partialwaves}).

Therefore we assume that the colliding particles are neutral and
spinless, and the Hamiltonian $h$ acts in the space spanned by only
two of its eigenstates,
\begin{equation}
\label{h_eigen_eq}
  h|n\rangle = E_n|n\rangle\ ,\qquad n=1,2\ ,
\end{equation}
i. e.
\begin{equation}
\label{h_spectral_exp}
  h= E_1|1\rangle\langle 1|+E_2|2\rangle\langle 2|\ .
\end{equation}
In its turn, this implies that the operators $H_0$ and $U$ are
$2\times2$ matrices in the subspace associated with the internal
degrees of freedom.  Therefore the total Hamiltonian (\ref{Hamiltonian})
taken in the coordinate representation and sandwiched between 
$\langle n|$ and $|n'\rangle$, has the following matrix representation
\begin{equation}
\label{Hmatrix}
  H=\left(
  \begin{array}{cc}
  \displaystyle
  -\frac{\hbar^2}{2\mu_1}\Delta_{\vec r}+U_{11}(\vec r)+
   E_1& U_{12}(\vec r)\\[3mm]
   U_{21}(\vec r) &    \displaystyle
  -\frac{\hbar^2}{2\mu_2}\Delta_{\vec r}+U_{22}(\vec r)+E_2
  \end{array}
  \right)\ ,
\end{equation}
where $\vec r$ is the relative coordinate and the subscripts label
the channels. A solution of the corresponding Schr\"odinger equation
\begin{equation}
\label{ Schreq}
    H\psi= E\psi
\end{equation}
is a column matrix
$$
     \psi=\left(
     \begin{array}{c}
     \psi_1(E,\vec r)\\[3mm]
     \psi_2(E,\vec r)
     \end{array}    
     \right)
$$
composed of the $\vec r$-dependent coefficients in the expansion
of the state vector over the channel (internal) states, i.e.
\begin{equation}
\label{expansion}
    \langle\vec r|\Psi\rangle = \psi_1(E,\vec r)|1\rangle+
    \psi_2(E,\vec r)|2\rangle\ .
\end{equation}
Continuing to keep unnecessary complications away, we assume that the
interaction potentials are spherically symmetric which means that the
orbital angular momentum $\ell$ associated with $\vec r$, is conserving
and hence is the same for the {\it in} and {\it out}
channels. Furthermore, for the matrix elements of the potential, we
assume that
\begin{equation}
\label{Vcondition}
     \int_0^\infty |U_{nn'}(r)|rdr<\infty\ ,
\end{equation}
i.e. that their possible singularity at $r=0$ is less than $1/r^2$
and they vanish at infinity faster than $1/r^2$.
Expanding the channel wave function in partial waves,
\begin{equation}
\label{partialwaves}
    \psi_n(E,\vec r)=\frac{u_n(E,r)}{r}Y_{\ell m}(\hat{\vec r})\ ,
\end{equation}
we obtain a set of coupled radial Schr\"odinger equations
\begin{equation}
\label{Schr_radial_eq}
   {\cal H}u=Vu\ ,
\end{equation}
where $u$ is a column matrix consisting of $u_1(E,r)$ and
$u_2(E,r)$, matrix elements of $V$ are
\begin{equation}
\label{reducedpotential}
    V_{nn'}(r)=\frac{2\mu_n}{\hbar^2}U_{nn'}(r)\ ,
\end{equation}
and
\begin{equation}
\label{Hradial}
     {\cal H}=
     \left(
     \begin{array}{cc}
     \displaystyle
     \frac{d^2}{dr^2}+k_1^2-\frac{\ell(\ell+1)}{r^2} & 
     0\\[3mm]
     0 &      
     \displaystyle
     \frac{d^2}{dr^2}+k_2^2-\frac{\ell(\ell+1)}{r^2}
     \end{array}
     \right)\ ,
\end{equation}
with the channel momenta $k_n$ defined as
\begin{equation}
\label{channelmomentum}
    k_n^2=\frac{2\mu_n}{\hbar^2}(E-E_n)\ .
\end{equation}
As is well known (from the general theory of ordinary differential
equations), a coupled set of $N$ equations of the type
(\ref{Schr_radial_eq}) has $2N$ linearly independent column solutions,
and only half of them are regular at $r=0$. Since in our case $N=2$,
we have two independent regular column-solutions that can be combined
in a square $2\times2$ matrix
\begin{equation}
\label{regsolmatrix}
     \phi(E,r)=\left(
     \begin{array}{ccc}
     \phi_{11}(E,r) && \phi_{12}(E,r)\\[3mm]
     \phi_{21}(E,r) && \phi_{22}(E,r)
     \end{array}
     \right)\ ,    
\end{equation}
where the first subscript labels the channels and the second is the
solution number. For the independent regular solutions, we use the
symbol $\phi$ to distinguish them from the physical solutions
$u$. By definition, they obey the same equation, namely,
\begin{equation}
\label{shcrmatrix}
     {\cal H}\phi=V\phi\ .
\end{equation}
Since all physical solutions $u(E,r)$ must be regular at $r=0$, anyone of
them is a linear combination of the regular columns, i.e.
\begin{equation}
\label{physsolmatrix}
     \left(\begin{array}{c}
     u_{1}(E,r)\\[3mm]
     u_{2}(E,r)
     \end{array}\right)
     =
     \left(\begin{array}{c}
     \phi_{11}(E,r)\\[3mm]
     \phi_{21}(E,r)
     \end{array}\right)c_1
     +
     \left(\begin{array}{c}
     \phi_{12}(E,r)\\[3mm]
     \phi_{22}(E,r)
     \end{array}\right)c_2\ ,
\end{equation}
or in a more compact form
\begin{equation}
\label{physsolmatrixcompact}
     \left(\begin{array}{c}
     u_{1}\\[3mm]
     u_{2}
     \end{array}\right)
     =
     \left(
     \begin{array}{cc}
     \phi_{11} & \phi_{12}\\[3mm]
     \phi_{21} & \phi_{22}
     \end{array}
     \right)
     \left(\begin{array}{c}
     c_1\\[3mm]
     c_2
     \end{array}\right)\ .
\end{equation}
The combination coefficients $c_1$ and $c_2$ should be chosen in such a way
that guarantees a given (physical) behavior of the solution when
$r\to\infty$.

\section{Transformation to first-order equations}
If the right hand side of Eq. (\ref{shcrmatrix}) were zero, the
equations of the set would be uncoupled (independent) and could
be solved analytically. Indeed, the Riccati-Hankel functions
$h_\ell^{(\pm)}(k_nr)$ are solutions of the equation
$$
    \left[\frac{d^2}{dr^2}+k_n^2-\frac{\ell(\ell+1)}{r^2}\right]
     h_\ell^{(\pm)}(k_nr)=0\ .
$$
Therefore the matrices
\begin{equation}
\label{INmatrix}
    W^{(\rm in)}_\ell(E,r)=
    \left(\begin{array}{cc}
    h_\ell^{(-)}(k_1r) & 0\\[3mm]
    0 & h_\ell^{(-)}(k_2r)
    \end{array}\right)
\end{equation}
and
\begin{equation}
\label{OUTmatrix}
    W^{(\rm out)}_\ell(E,r)=
    \left(\begin{array}{cc}
    h_\ell^{(+)}(k_1r) & 0\\[3mm]
    0 & h_\ell^{(+)}(k_2r)
    \end{array}\right)
\end{equation}
both solve Eq. (\ref{shcrmatrix}) when its right hand side is zero, i.e.
\begin{equation}
\label{HWzero}
    {\cal H}W_\ell^{(\rm in/out)}(E,r)=0\ .
\end{equation}
These matrices represent the incoming and outgoing spherical waves
which form the asymptotics of the solution at large distances when the
potential vanishes. In order to guarantee correct asymptotic behavior
of the solutions (\ref{regsolmatrix}) of Eq. (\ref{shcrmatrix}), let
us look for them in the following form
\begin{equation}
\label{ansatzMatrix}
     \phi(E,r)= W^{(\rm in)}_\ell(E,r){\cal F}^{(\rm in)}(E,r)+ 
     W^{(\rm out)}_\ell(E,r){\cal F}^{(\rm out)}(E,r)\ ,
\end{equation}
where ${\cal F}^{(\rm in/out)}(E,r)$ are new unknown matrix
functions. In the theory of ordinary differential equations, this way
of finding solution is known as the {\it variation parameters method}.

Since we replaced one unknown matrix $\phi$ with two unknown matrices
${\cal F}^{(\rm in/out)}$, they cannot be independent.  We therefore can
impose one arbitrary condition that relates them to each other. As such
condition, we choose the following equation
\begin{equation}
\label{Lagrange}
     W^{(\rm in)}_\ell(E,r)\frac{d}{dr}{\cal F}^{(\rm in)}(E,r)+ 
     W^{(\rm out)}_\ell(E,r)\frac{d}{dr}{\cal F}^{(\rm out)}(E,r)=0\ ,
\end{equation}
which is standard in the variation parameters method and is called the
{\it Lagrange} condition. It implies that, when calculating first
derivative of $\phi(E,r)$, we only need to differentiate the spherical 
waves,
\begin{equation}
\label{phiFirstDerivative}
   \frac{d}{dr}\phi(E,r)=W'^{\,(\rm in)}_\ell(E,r){\cal F}^{(\rm in)}(E,r)+ 
     W'^{\,(\rm out)}_\ell(E,r){\cal F}^{(\rm out)}(E,r)\ .
\end{equation}
Let us substitute the ansatz (\ref{ansatzMatrix}) into
Eq. (\ref{shcrmatrix}) and thus obtain the corresponding equations
that determine ${\cal F}^{(\rm in)}$ and ${\cal F}^{(\rm out)}$.
The second derivative of $\phi(E,r)$ is
$$
   \frac{d^2}{dr^2}\phi=
   W''^{\,(\rm in)}_\ell{\cal F}^{(\rm in)}+ 
   W''^{\,(\rm out)}_\ell{\cal F}^{(\rm out)}+
   W'^{\,(\rm in)}_\ell{\cal F}'^{\,(\rm in)}+ 
   W'^{\,(\rm out)}_\ell{\cal F}'^{\,(\rm out)}\ .
$$    
This means that
$$
   {\cal H}\phi=\left({\cal H}W^{(\rm in)}_\ell\right){\cal F}^{(\rm in)}+
   \left({\cal H}W^{(\rm in)}_\ell\right){\cal F}^{(\rm in)}+
   W'^{\,(\rm in)}_\ell{\cal F}'^{\,(\rm in)}+ 
   W'^{\,(\rm out)}_\ell{\cal F}'^{\,(\rm out)}\ ,  
$$
where the first two terms disappear in accordance with Eq. (\ref{HWzero}).
Therefore Eq. (\ref{shcrmatrix}) takes the form
\begin{equation}
\label{intermediate1}
   W'^{\,(\rm in)}_\ell{\cal F}'^{\,(\rm in)}+ 
   W'^{\,(\rm out)}_\ell{\cal F}'^{\,(\rm out)}=
   V\left[W^{(\rm in)}_\ell{\cal F}^{(\rm in)}+ 
   W^{(\rm out)}_\ell{\cal F}^{(\rm out)}\right]\ .
\end{equation}
From the Lagrange condition (\ref{Lagrange}), it follows that
\begin{equation}
\label{intermediate2}
   {\cal F}'^{\,(\rm out)}=-\left[W^{(\rm out)}_\ell\right]^{-1}
   W^{(\rm in)}_\ell{\cal F}'^{\,(\rm in)}\ .
\end{equation}
It should be noted that since matrices $W^{(\rm in/out)}_\ell$ and their
derivatives all are diagonal, they commute with each other.
Substituting Eq. (\ref{intermediate2}) into  Eq. (\ref{intermediate1})
and using this commutative property, we obtain
\begin{equation}
\label{FinEq1}
   {\cal F}'^{\,(\rm in)}=
   \left[
   W^{(\rm out)}_\ell W'^{(\,\rm in)}_\ell-W'^{(\,\rm out)}_\ell
   W^{(\rm in)}_\ell   \right]^{-1}W^{(\rm out)}_\ell
   V\left[W^{(\rm in)}_\ell{\cal F}^{(\rm in)}+ 
   W^{(\rm out)}_\ell{\cal F}^{(\rm out)}\right]\ .
\end{equation}
Similarly, from the Lagrange condition (\ref{Lagrange}), it follows that
\begin{equation}
\label{intermediate3}
   {\cal F}'^{\,(\rm in)}=-\left[W^{(\rm in)}_\ell\right]^{-1}
   W^{(\rm out)}_\ell{\cal F}'^{\,(\rm out)}
\end{equation}
and therefore Eq. (\ref{intermediate1}) gives
\begin{equation}
\label{FoutEq1}
   {\cal F}'^{\,(\rm out)}=
   \left[
   W^{(\rm in)}_\ell W'^{(\,\rm out)}_\ell-W'^{(\,\rm in)}_\ell 
   W^{(\rm out)}_\ell   \right]^{-1}W^{(\rm in)}_\ell
   V\left[W^{(\rm in)}_\ell{\cal F}^{(\rm in)}+ 
   W^{(\rm out)}_\ell{\cal F}^{(\rm out)}\right]\ .
\end{equation}
The first factors in Eqs. (\ref{FinEq1}) and (\ref{FoutEq1}) differ
only by the sign and actually are equal to the inverted Wronskian of the
incoming and outgoing spherical waves, which can be found in explicit
form. Indeed, knowing the Wronskian of the Riccati-Hankel functions,
\begin{equation}
\label{WronskianHpm}
   h_\ell^{(-)}(kr)\frac{d}{dr}h_\ell^{(+)}(kr)-
   \frac{d}{dr}h_\ell^{(-)}(kr)h_\ell^{(+)}(kr)=2ik\ ,
\end{equation}
we obtain
\begin{equation}
\label{WronskianWinout}
   W^{(\rm in)}_\ell W'^{(\,\rm out)}_\ell-W'^{(\,\rm in)}_\ell
   W^{(\rm out)}_\ell   =2iK\ ,
\end{equation}
where
\begin{equation}
\label{MomentumMatrix}
   K=
   \left(\begin{array}{cc}
   k_1 & 0\\[3mm]
   0 & k_2
   \end{array}\right)\ .   
\end{equation}
Therefore the Schr\"odinger equation (\ref{shcrmatrix}) is equivalent to
the following set of coupled equations of the first order
\begin{equation}
\label{FinFoutMatrixEq}
   \left\{
   \begin{array}{lcr}
   \displaystyle
   \frac{d}{dr}{\cal F}^{(\rm in)} &=&
   \displaystyle
   -\frac{1}{2i}K^{-1}W^{(\rm out)}_\ell
   V\left[W^{(\rm in)}_\ell{\cal F}^{(\rm in)}+ 
   W^{(\rm out)}_\ell{\cal F}^{(\rm out)}\right]\\[5mm]
   \displaystyle
   \frac{d}{dr}{\cal F}^{(\rm out)} &=&
   \displaystyle
   \frac{1}{2i}K^{-1}W^{(\rm in)}_\ell
   V\left[W^{(\rm in)}_\ell{\cal F}^{(\rm in)}+ 
   W^{(\rm out)}_\ell{\cal F}^{(\rm out)}\right]
   \end{array}   
   \right.\ .
\end{equation}
Boundary conditions at $r=0$ for equations (\ref{FinFoutMatrixEq})
follow from the requirement that the solution (\ref{ansatzMatrix})
must be regular.  This can only be achieved if ${\cal F}^{(\rm
in)}(E,r)$ and ${\cal F}^{(\rm out)}(E,r)$ become identical when $r\to
0$ because in such a case the singular parts of $h_\ell^{(-)}(kr)$
and $h_\ell^{(+)}(kr)$ cancel each other, i.e.
\begin{equation}
\label{HpmF}
   h_\ell^{(-)}(kr)+h_\ell^{(+)}(kr)=2j_\ell(kr)\ .
\end{equation}
Here the Riccati-Bessel function $j_\ell$ is regular, namely,
\begin{equation}
\label{CoulombRegularZero}
   j_\ell(kr)\ \mathop{\longrightarrow}_{r\to0}\ 
   \frac{\sqrt{\pi}}{2^{\ell+1}\Gamma(\ell+3/2)}(kr)^{\ell+1}\ .
\end{equation}
Matrices ${\cal F}^{(\rm in)}(E,r)$ and ${\cal F}^{(\rm out)}(E,r)$
must become diagonal near $r=0$ in order to guarantee linear
independence of the columns of $\phi(E,r)$.  Thus, the
following boundary conditions
\begin{equation}
\label{regularBC}
   \lim_{r\to0}{\cal F}_{nn'}^{(\rm in/out)}(E,r)=\delta_{nn'}
\end{equation}
are appropriate since we have the freedom to choose the normalization
of the solution. 

\section{Jost matrices}

The differential equations (\ref{FinFoutMatrixEq}) can be numerically
solved from the origin to a sufficiently far point $r=R$ where the
potential vanishes (causing the right-hand sides of the equations to
disappear) and therefore ${\cal F}^{(\rm in/out)}(E,r)$ become
constant. These constants
\begin{equation}
\label{fInDefinition}
     f^{(\rm in)}(E)=\lim_{r\to\infty}{\cal F}^{(\rm in)}(E,r)
\end{equation}
and
\begin{equation}
\label{fOutDefinition}
     f^{(\rm out)}(E)=\lim_{r\to\infty}{\cal F}^{(\rm out)}(E,r)
\end{equation}
are the Jost matrices that determine asymptotic behavior of the 
fundamental system of regular solutions
\begin{equation}
\label{AsymptoticMatrix}
     \phi(E,r)\ \mathop{\longrightarrow}_{r\to\infty}\ 
     W^{(\rm in)}_\ell(E,r)f^{(\rm in)}(E)+ 
     W^{(\rm out)}_\ell(E,r)f^{(\rm out)}(E)\ .
\end{equation}
It is worthwhile to mention that we use here the notation $f^{(\rm
in/out)}(E)$ which is different from the traditional notation such as
$f^{(\pm)}(\pm k)$. There are two reasons for this. First of all, we do not
fix the normalization of the regular solution $\phi(E,r)$.  As a
result, both $f^{(\rm in)}(E)$ and $f^{(\rm out)}(E)$ can have an
arbitrary common factor. We are not concerned with this factor because
no observable quantity depends on it. In contrast to a majority of other
studies, we leave the normalization of the Jost matrices free,
and therefore need a notation that is different from the traditional.
The second reason is that the superscripts ({\it in}) and ({\it out})
are unambiguous and thus we avoid possible confusion caused by the
existence of notations with opposite signs for the same Jost matrices.

The regular solution $\phi(E,r)$ consists of two linearly independent 
columns which constitute the fundamental system of solutions. Any other
solution is a linear combination of these independent columns.
The physical wave function (\ref{physsolmatrix}) is one of such
combinations,
\begin{equation}
\label{PhysicalSolution}
      u(E,r)=\phi(E,r)c\ ,
\end{equation}
where
$$
    u=\left(\begin{array}{c}u_1\\u_2\end{array}\right)
    \qquad \mbox{and} \qquad
    c=\left(\begin{array}{c}c_1\\c_2\end{array}\right)\ .
$$
At large distances, we therefore have
\begin{equation}
\label{PhysWfc}
     u(E,r)\ \mathop{\longrightarrow}_{r\to\infty}\ 
     W^{(\rm in)}_\ell(E,r)f^{(\rm in)}(E)
     \left(\begin{array}{c}c_1\\c_2\end{array}\right)
     + 
     W^{(\rm out)}_\ell(E,r)f^{(\rm out)}(E)
     \left(\begin{array}{c}c_1\\c_2\end{array}\right)\ .
\end{equation} 

\subsection{Spectral points}
There are certain discrete points in the complex $E$-plane 
 (bound, resonant, and virtual states), at
which the physical wave function has only outgoing waves in its
asymptotic behavior. At these so called {\it spectral points} the 
combination coefficients $c_1$ and 
$c_2$ are such that
$$
     \left(\begin{array}{cc}
     f^{(\rm in)}_{11}(E) & f^{(\rm in)}_{12}(E)\\[3mm]
     f^{(\rm in)}_{21}(E) & f^{(\rm in)}_{22}(E)\end{array}\right)
     \left(\begin{array}{c}c_1\\[3mm]
     c_2\end{array}\right)
     =0\ .
$$
This homogeneous equation has a non-trivial solution if and only if
\begin{equation}
\label{detJostMatrixZero}
     \det f^{(\rm in)}(E)=0\ .
\end{equation}
As is seen, the distribution of the spectral points in the complex
$E$-plane (determined by this equation) does not, as was mentioned earlier,
 depend on the
normalization of the Jost matrix.

\subsection{Scattering}
For a scattering state, the wave function has both incoming and
outgoing waves at large $r$.  If $A_n$ and $B_n$ are the amplitudes
of the incoming and outgoing (scattered) waves in the $n$-th channel,
then the scattering boundary condition (at $r\to\infty$) reads
\begin{eqnarray}
\label{AscattBC1}
   u(E,r)
   &\mathbin{\mathop{\longrightarrow}\limits_{r\to\infty}}&
   \left(
   \begin{array}{c}
   h^{(-)}_\ell(k_1r)A_1+
   h^{(+)}_\ell(k_1r)B_1\\[3mm]
   h^{(-)}_\ell(k_2r)A_2+
   h^{(+)}_\ell(k_2r)B_2
   \end{array}
   \right)\\[5mm]
\label{AscattBC2}
   &=&
   W^{(\rm in)}_\ell(E,r)\left(
   \begin{array}{c}
   A_1\\[3mm] A_2 \end{array}\right)
   +
   W^{(\rm out)}_\ell(E,r)\left(
   \begin{array}{c}
   B_1\\[3mm] B_2 \end{array}\right)\ .
\end{eqnarray}
Comparing Eqs. (\ref{AscattBC2}) and (\ref{PhysWfc}), we see that
\begin{equation}
\label{AscattABc1c2}
   \left(\begin{array}{c}
   A_1 \\ A_2
   \end{array}\right)
   = f^{(\rm in)}(E)
   \left(\begin{array}{c}
   c_1 \\ c_2
   \end{array}\right)
\end{equation}
and
\begin{equation}
\label{AscattApBpc1c2}
   \left(\begin{array}{c}
   B_1 \\ B_2
   \end{array}\right)
   = f^{(\rm out)}(E)
   \left(\begin{array}{c}
   c_1 \\ c_2
   \end{array}\right)\ .
\end{equation}
This implies that the amplitudes of the incident and scattered waves are
related by a $2\times2$-matrix , $S(E)$ as
\begin{equation}
\label{ASmatr}
   \left(\begin{array}{c}
   B_1 \\ B_2
   \end{array}\right)
   = S(E)
   \left(\begin{array}{c}
   A_1 \\ A_2
   \end{array}\right)\ .
\end{equation}
This $S$-matrix
\begin{equation}
\label{ASmatrdef}
   S(E)
   = f^{(\rm out)}(E)\left[f^{(\rm in)}(E)\right]^{-1}
\end{equation}
does not depend on the choice of $A_n$. We can therefore clarify the
physical meaning of its matrix elements by considering special cases
with simple choices of $A_1$ and $A_2$.  If the incident
 wave is purely channel 1, i.e.  $A_1\ne0$ and $A_2=0$, then
$R_{11}=B_1/A_1$ and $T_{12}=B_2/A_1$ are the amplitudes of the
probability that the incoming wave is returning in the channel 1 and
is transmitted into channel 2, respectively. Similarly, if $A_1=0$
and $A_2\ne0$ then $R_{22}=B_2/A_2$ and $T_{21}=B_1/A_2$ are the
amplitudes of the probability that the incoming wave is returning in
the channel 2 and is transmitted into channel 1, respectively.
Substituting $A_1=0$ or $A_2=0$ into the equation
\begin{equation}
\label{ASmatrmeaning}
	       \left(
	       \begin{array}{c}
	       B_1\\ B_2
	       \end{array}
	       \right)=\left(
	       \begin{array}{c}
	       S_{11}A_1+S_{12}A_2\\ S_{21}A_1+S_{22}A_2
	       \end{array}
	       \right)
\end{equation}
we see that the $S$--matrix consists of these transmission and reflection
amplitudes, namely,
\begin{equation}
\label{AStr}
	       S(E)=
	       \left(
	       \begin{array}{ccc}
	       R_{11}(E) & ,& T_{21}(E)\\
	       T_{12}(E) & ,& R_{22}(E)
	       \end{array}
	       \right)\ .
\end{equation}
In general the energy $E$ is complex
and only for real energies the quantities $R_{nn}$ and $T_{nn'}$ have
the simple physical meaning. 

Also note that the $S$-matrix and therefore the
observable transmission and reflection amplitudes do not depend on the
normalization of the Jost matrices. Indeed, any common factor of
$f^{(\rm in)}$ and $f^{(\rm out)}$ cancels out in
Eq. (\ref{ASmatrdef}) which supports our suggestion to move away
from their fixed normalization.

\subsection{Partial widths}
At the energy points corresponding to bound
states, and at complex energies in the resonance region (fourth
quadrant) of the $E$-plane the elements of the $S$-matrix have special
properties as well. From Eqs. (\ref{ASmatrdef})
and (\ref{detJostMatrixZero}), we see that at every bound state and
resonance 
\begin{equation}
\label{AEres}
    E_{\rm res}=E_{\rm r}-i\frac{\Gamma}{2}
\end{equation}
the $S$-matrix has a pole. If the energy is close to 
$E_{\rm res}$ then 
\begin{equation}
\label{ASnearEres}
    S_{nn'}(E)
    \mathbin{\mathop{\approx}\limits_{E\to E_{\rm res}}}
    {\rm const}\left(1-\frac{i\sqrt{\Gamma_n\Gamma_{n'}}}
    {E-E_{\rm r}+i{\Gamma}/{2}}\right)\ ,
\end{equation}
where $\Gamma_1$ and $\Gamma_2$ are the partial widths such that
together they form the total width 
\begin{equation}
\label{AGamma}
    \Gamma=\Gamma_1+\Gamma_2
\end{equation}
of the resonance, and the ratios $\Gamma_1/\Gamma$ and
$\Gamma_2/\Gamma$ are the probabilities that the resonance will decay
through (or can be excited from) the first or second channels,
respectively.  
As it follows from Eq. (\ref{ASnearEres}), the partial widths
$\Gamma_n$ can be found as the limits
$$
       \Gamma_n=\lim_{E\to E_{\rm res}}(E-E_{\rm res})S_{nn}(E)\ .
$$ 
However, in numerical calculations, finding a limit of a singular
functions is not an easy task. A way to avoid this difficulty was
suggested by Masui et al. in Ref. \cite{Masui}.
From Eq. (\ref{ASnearEres}) it is clear that
\begin{equation}
\label{AGratio}
    \mathbin{\mathop{\lim}\limits_{E\to E_{\rm res}}}
    \left|
    \frac{S_{11}(E)}{S_{22}(E)}
    \right|
    =
    \frac{\Gamma_1}{\Gamma_2}\ .
\end{equation}
Therefore the partial widths can be found with the help of
Eqs. (\ref{AGamma}) and (\ref{AGratio}), if $\Gamma$ is known and if
there is a procedure for the $S$-matrix calculation at complex
energies.

We can significantly simplify the task of finding the limit
(\ref{AGratio}) by using Eq. (\ref{ASmatrdef}).  Indeed, both $S_{11}$
and $S_{22}$ have singularities at $E_{\rm res}$ because of vanishing
determinant of $f^{(\rm in)}$, which cancels out in their
ratio. Making explicit inversion of the $2\times2$ matrix $f^{(\rm
in)}$ and using Eq. (\ref{ASmatrdef}), we obtain
\begin{equation}
\label{GGratioExplicit}
     \frac{\Gamma_1}{\Gamma_2}=\left|
     \frac{f^{(\rm out)}_{11}f^{(\rm in)}_{22}
          -f^{(\rm out)}_{12}f^{(\rm in)}_{21}}
          {f^{(\rm out)}_{22}f^{(\rm in)}_{11}
          -f^{(\rm out)}_{21}f^{(\rm in)}_{12}}\right|_{E=E_{\rm res}}\ ,
\end{equation}
where no singularities are present.

\section{Complex energies}
The differential equations (\ref{FinFoutMatrixEq}) enable us to obtain
a complete solution of the two-channel (generally, $N$-channel)
problem at any energy of physical interest. Their advantage over the
corresponding Schr\"odinger equation becomes especially evident when
we consider complex values of the energy.

\subsection{Riemann surface}
Since the functions ${\cal F}{({\rm in/out})}(E,r)$, and therefore the Jost
matrices , depend on the energy $E$ via the
channel momenta $k_1=\sqrt{(2\mu_1/\hbar^2)(E-E_1)}$ and
 $k_2=\sqrt{(2\mu_2/\hbar^2)(E-E_2)}$, there are two square-root
branching points for these matrices in the energy plane :
$E=E_1$ and $E=E_2$. This means that if we make
two full circles around either of these points then matrix elements of
$f^{({\rm in/out})}(E)$ return to the same values from which the circling
was started (one full circle is not enough). This is because the
channel momentum
\begin{equation}
\label{Amomentum}
       k_n=\sqrt{(2\mu_n/\hbar^2)|E-E_n| e^{i\chi_n}}
       =\sqrt{(2\mu_n/\hbar^2)|E-E_n|}e^{i\chi_n/2}
\end{equation}
comes to its initial value when $\chi_n \to \chi_n+4\pi$, where 
$|E-E_n|$ and $\chi_n$ are the polar coordinates of the point
$E$ on the energy-plane relative to the branching point
$E_n$, i.e.
\begin{equation}
\label{Bpolar}
		E=E_n+|E-E_n|e^{i\chi_n}\ .
\end{equation}
In other words, elements of the Jost matrix have two different
values at each point $E$ on this circle.

\begin{figure}[htp]
\centerline{\epsfig{file=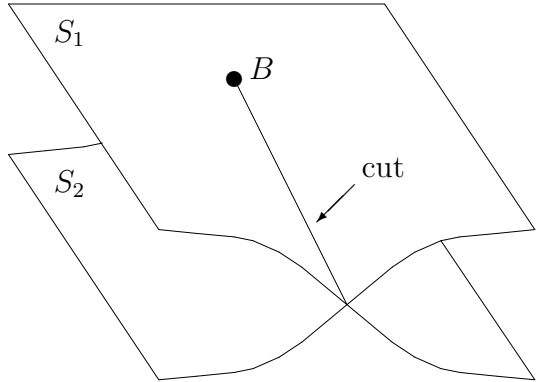}}
\caption{\sf
Fragments of the physical ($S_1$) and unphysical ($S_2$) sheets of the
complex-energy Riemann surface around a branching point ($B$) that 
corresponds to a threshold energy ($E_2$ or $E_1$).
Transition from $S_1$ to $S_2$ and back is possible through the unitary
cut running from $B$ to infinity.
}
\label{fig.sheets}
\end{figure}

In order to make the Jost matrix a single-valued function of $E$, we
can assume (as is usual in the complex analysis) that the complex
energy forms the so-called Riemann surface consisting of several
parallel sheets. When doing the first circle around a branching point,
we are moving on the first sheet and then continue on the second one
until coming back to the first sheet after completing the full two
circles.  Such continuous transition from one sheet to another is
possible if we make a cut from the branching point to infinity, and
connect opposite rims of the cuts on the two sheets (see
Fig. \ref{fig.sheets}).

As is usual in the scattering theory, we make straight line cuts from
both branching points to infinity along the positive real axis.  Each
of the two sheets related by the first branching point, is further
branched at the second branching point. Therefore the full Riemann
surface consists of four parallel sheets. We can reach any of these
sheets by making an appropriate number of circles around the first and
the second branching points.

The physical energy (at which the scattering takes place) is on the
positive real axis. We choose the cuts and their interconnections in
such a way that the physical scattering energies lie on the upper rims
of the both cuts.  Starting from these physical energies and moving in
the anti-clockwise direction around all branching points, we cover the
so-called physical sheet of the energy plane. According to
Eq. (\ref{Amomentum}) the channel momenta corresponding to this sheet,
have positive imaginary parts. The bound states are also on the
physical sheet (this is necessary to guarantee the exponential
attenuation of their wave functions). The resonances, however,
correspond to zeros of the Jost-matrix determinant at the momenta with
negative imaginary parts  and therefore lie on the unphysical sheet of the
Riemann surface. The resonances that are able to decay into both
channels must have ${\rm Im\,}(k_1)<0$ and ${\rm
Im\,}(k_2)<0$. This means that they lie on the sheet which is
unphysical with respect to the both branching points.

In practical calculations, we can choose point on one of the four
Riemann sheets by selecting appropriate signs in front of the square
roots for $k_1$ and $k_2$, i.e. appropriate signs of their imaginary
parts.  These signs must be such that the imaginary part of the
momentum is positive for a closed channel and negative for an open
one. This corresponds to the choice between the physical and
unphysical sheets of the Riemann $E$-surface and ensures correct
asymptotic behavior of the wave function.

\subsection{Complex rotation}
Similarly to the single-channel case (see Refs.
\cite{nuovocim,exactmethod}), the scheme described in previous
sections, can be easily implemented only for bound and scattering
states. In the resonance domain of the complex $E$-plane (below the
real axis) the Riccati-Hankel function $h_\ell^{(+)}(kr)$, as can be
seen from its asymptotics
\begin{equation}
\label{HpmAsymptotics}
    h_\ell^{(\pm)}(z)\,\,\mathop{\longrightarrow}_{|z|\to\infty}\,\,
    \mp i\exp\left\{\pm i[z-\ell\pi/2]\right\}\ ,
\end{equation}
diverges when $r\to\infty$.  As a
result the right hand side of the first matrix equation of the set
(\ref{FinFoutMatrixEq}) diverges and hence the limit
(\ref{fInDefinition}) cannot be calculated. It should be emphasized
that this fact does not mean that this limit does not exist. It do
exist, but simply moving along the real $r$-axis, we cannot reach it.
It is easy to see, that
\begin{equation}
\label{HankelPlusLargeRzero}
    \mbox{if}\ \ {\rm Im\,}kr>0\ \Longrightarrow
    h_\ell^{(+)}(kr)\ \mathop{\longrightarrow}_{|kr|\to0}\ 0\ .
\end{equation}
When ${\rm Im\,}kr=0$ this function remains finite (oscillates) at
large $r$.  The condition ${\rm Im\,}k_nr>0$ for asymptotic vanishing
of the function $h_\ell^{(+)}(\eta_n,k_nr)$ involves the imaginary
part of the product $k_nr$ but not of the momentum alone. This offers
an elegant way to extend the domain of the $E$-plane where the limit
(\ref{fInDefinition}) can be calculated, to practically whole
$E$-plane.  Indeed, if, for example, ${\rm Im\,}k_nr$ is negative we
can always make it positive by using complex values of $r$.  This of
course requires that the potential is defined for complex $r$ and
tends to zero when $|r|\to\infty$ at least in certain sector of the
complex $r$-plane. We assume that it vanishes faster than $r^{-2}$
when $r\to\infty$ along any line
\begin{equation}
\label{rot}
	  r=z\exp(i\theta)\ , \qquad z\ge 0\ ,
\end{equation}
for the rotation angle $\theta$ in the interval
$0\le|\theta|\le\theta_{\rm max}<{\pi}/{2}$. Then, firstly, according
to the existence theorem, the solutions $\phi_{nn'}(E,r)$ of the
Schr\"odinger equation are holomorphic functions of $r$ within the
corresponding domain of the $r$-plane and, secondly, they have the
asymptotic behavior of the type (\ref{AsymptoticMatrix}) along any
line (\ref{rot}).  Moreover, the coefficients $f^{(\rm in/out)}(E)$
in such asymptotics are the same for all choices of the
rotation angle because they do not depend on $r$.

Similarly to Eq. (\ref{HankelPlusLargeRzero}), we have
\begin{equation}
\label{HankelMinusLargeRzero}
    \mbox{if}\ \ {\rm Im\,}kr<0\ \Longrightarrow
    h_\ell^{(-)}(kr)\ \mathop{\longrightarrow}_{|kr|\to0}\ 0\ . 
\end{equation}

Therefore, the limiting values ${\cal F}^{\rm (in/out)}(E,\infty)$ can
be found as the corresponding coefficients in the asymptotics of the
functions ${\cal F}^{\rm (in/out)}(E,ze^{i\theta})$.  It can be shown
that with ${\rm Im\,}kr\ge 0$ the right hand side of the first
equation of the set (\ref{FinFoutMatrixEq}) at large distances tends
to zero, which means vanishing of the derivative on the left hand
side, i.e. that ${\cal F}^{(\rm in)}$ becomes constant (reaches its
limit).  Similarly, it can be shown that the limit
(\ref{fOutDefinition}) can be reached if ${\rm Im\,}kr\le 0$. Both
limits, i.e.  $f^{(\rm in)}(E)$ and $f^{(\rm out)}(E)$ can be found
simultaneously only when ${\rm Im\,}kr= 0$ and at all spectral points.
The line  ${\rm Im\,}kr= 0$ therefore serves as the dividing line that
separates two domains of the complex energy plane. The Jost matrix
$f^{(\rm in)}(E)$ can be calculated above this line, while 
$f^{(\rm out)}(E)$ below it.

\begin{figure}[htp]
\centerline{\epsfig{file=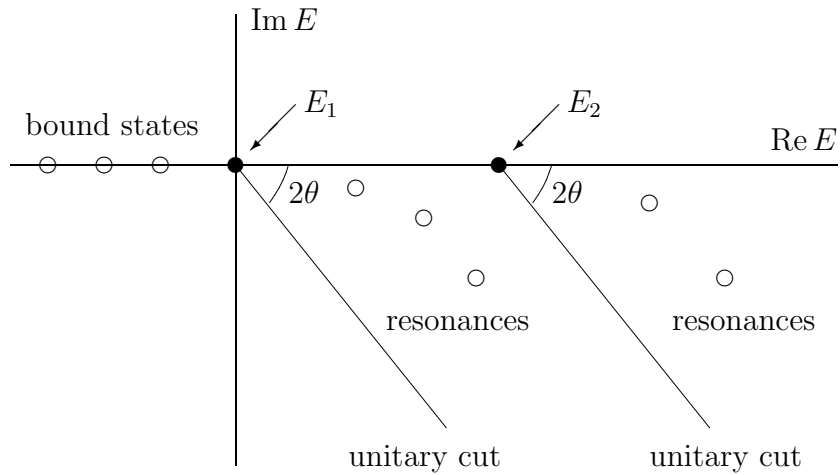}}
\caption{\sf 
Typical distribution of the bound states and resonances i.e. spectral 
points (open circles) in the complex energy-plane.  It is assumed
that $E_1<E_2$ and the energy is measured relative
to the lowest threshold $E_1$.    
The unitary cuts going from the branching points to infinity are also shown.
Because of the complex rotation, these cuts are turned into the
unphysical sheet by the angle $2\theta$.
}
\label{fig.spectralpoints}
\end{figure}

By considering complex $r$, we actually do the analytic continuation
of $f^{(\rm in/out)}(E)$ across the dividing lines (the unitary cuts)
to the domains where Eqs. (\ref{FinFoutMatrixEq}) do not give finite
values for these matrices.  The dividing lines can be
turned down wards to expose the resonance spectral points, by rotating
$r$ as given by Eq. (\ref{rot}). Indeed, if $\chi_n$ is the polar
angle parametrizing the position of a point on the $E$-plane relative
to the branching point $E_n$, then by choosing large enough
$\theta$ we can make ${\rm Im\,}k_nr$,
\begin{equation}
\label{imkr}
	{\rm Im\,}k_nr=
        {\rm Im\,}\left(|k_n|ze^{i(\theta+\chi_n/2)}\right)=
	|k_l|z\sin(\theta+\chi_n/2)\, ,
\end{equation}
positive even when $\chi_n$ is negative (when the point $E$ is below
the cut, i.e. on the unphysical sheet) and vice versa.  From the last
equation is clear that when $\theta > 0$ both dividing lines are turned
down by $2\theta$ (see Fig. \ref{fig.spectralpoints}).

The fact that $f^{(\rm in)}(E)$ and $f^{(\rm out)}(E)$ can be found
simultaneously only on the lines corresponding to the unitary cuts and
at the spectral points, does not pose a problem. Indeed, for bound
states ${\rm Im\,}kr>0$ and we can locate them using
Eq. (\ref{detJostMatrixZero}) without the rotation. As soon as a bound
state has been located at a spectral point $E_b$, we can calculate
both ${\cal F}^{(\pm)}(E_b,r)$ at this point and thus obtain the wave
function.  The scattering takes place exactly on the real axis that
coincides with the unitary cut where both $f^{(\rm in)}(E)$ and
$f^{(\rm out)}(E)$ can be calculated without rotation. To locate
resonances, we turn the cuts down opening the unphysical energy
sheets, and again use Eq. (\ref{detJostMatrixZero}) which involves
only $f^{(\rm in)}(E)$.  If we need to obtain the partial widths
(which requires the knowledge of the $f^{(\rm out)}(E)$ as well), we
have to repeat the integration of the differential equations at the
resonance energy with such rotation angle that the resonance in
question is below the cut.

It should be also mentioned that the border separating the two domains
of the complex $E$-plane is a line only in the case of potentials
slowly vanishing at infinity (for example, as $1/r^n$). If, however,
the potential decays exponentially, then $f^{(\rm in)}(E)$ can be
found not only above the dividing line but also within a band below
this line and $f^{(\rm out)}(E)$ within a symmetrical band above the
line (see Ref. \cite{partialwaves}).  The faster the potential decays
the wider this band is.

\subsection{Integration path}
As was shown above, when solving the differential equations
(\ref{FinFoutMatrixEq}), we have to move from the origin to infinity
(actually to a large $|r|=R$) along the ray (\ref{rot}) with generally
nonzero $\theta$.  In numerical calculations, however, this
straightforward approach is not always practical. As was mentioned
earlier, in the immediate vicinity of the point $r=0$, the
singularities of the Riccati-Hankel functions cancel each other in
accordance with Eq. (\ref{HpmF}). This may make
Eqs. (\ref{FinFoutMatrixEq}) numerically unstable near $r=0$. Such an
instability is further aggravated when the angular momentum $\ell$
becomes large.

\begin{figure}[htp]
\centerline{\epsfig{file=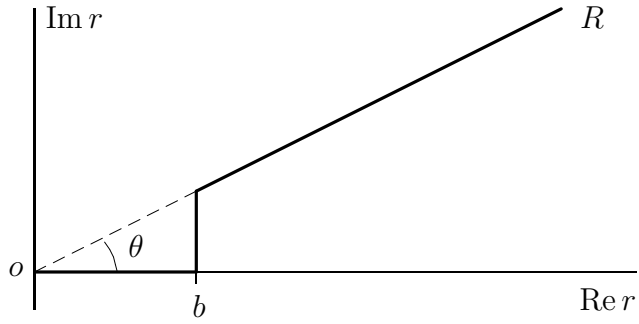}}
\caption{ 
The integration path for solving Eqs.
(\ref{ABMatrixEq}) on the segment $ob$, and the Jost function
equations (\ref{FinFoutMatrixEq}) on $bR$.  
}
\label{fig.IntegrationPath}
\end{figure}

This problem can be easily circumvented in two ways.  One way is
to solve the matrix Schr\"odinger equation (\ref{shcrmatrix})
on an interval from a sufficiently small $r=r_{\rm min}$ to an
intermediate point $r=b$.  Having thus calculated the matrices
$\phi(E,b)$ and $\phi'(E,b)$, we can obtain the values
${\cal F}^{(\rm in/out)}(E,b)$ using the relations
$$
     {\cal F^{(\rm in/out)}}(E,r) =
     \pm\frac{1}{2i}K^{-1}\left[
     W'^{(\rm out/in)}_J(E,r)\phi(E,r)-
     W_J^{(\rm out/in)}(E,r)\phi'(E,r)\right]\ ,
$$ 
which follow from Eqs. (\ref{ansatzMatrix}) and
(\ref{phiFirstDerivative}). Then we can turn to the ray (\ref{rot})
along the path shown in Fig. \ref{fig.IntegrationPath}, and safely
proceed with Eqs. (\ref{FinFoutMatrixEq}) to a sufficiently large
$z=R$, which will serve as the ``infinity''.

Another way is to rearrange Eqs. (\ref{FinFoutMatrixEq}) on the
interval $[r_{\rm min},b]$ in order to avoid the singularity
cancellations.  Since $(h^{(+)}_\ell+h^{(-)}_\ell)/2=j_\ell$ and
$(h^{(+)}_\ell-h^{(-)}_\ell)/2i=n_\ell$, where $n_\ell$ is the
Riccati-Neumann function (which is singular at $r=0$), we may
introduce a new pair of matrices
\begin{equation}
\label{Amatrix}
   {\cal A}(E,r)=
   {\cal F}^{(\rm in)}(E,r)+{\cal F}^{(\rm out)}(E,r)
\end{equation}
and
\begin{equation}
\label{Bmatrix}
   {\cal B}(E,r)=i\left[
   {\cal F}^{(\rm in)}(E,r)-{\cal F}^{(\rm out)}(E,r)\right]\ ,
\end{equation}
which transform the ansatz (\ref{ansatzMatrix}) into the form
\begin{equation}
\label{ABansatzMatrix}
     \phi(E,r)= {\cal J}_\ell(E,r){\cal A}(E,r)-
     {\cal N}_\ell(E,r){\cal B}(E,r)\ ,
\end{equation}
where
\begin{equation}
\label{Jmatrix}
    {\cal J}_\ell(E,r)=
    \left(\begin{array}{cc}
    j_\ell(k_1r) & 0\\[3mm]
    0 & j_\ell(k_2r)
    \end{array}\right)
\end{equation}
and
\begin{equation}
\label{Nmatrix}
    {\cal N}_\ell(E,r)=
    \left(\begin{array}{cc}
    n_\ell(k_1r) & 0\\[3mm]
    0 & n_\ell(k_2r)
    \end{array}\right)\ .
\end{equation}
The corresponding linear combination of Eqs. (\ref{FinFoutMatrixEq})
gives their alternative form
\begin{equation}
\label{ABMatrixEq}
   \left\{
   \begin{array}{lcr}
   \displaystyle
   \frac{d}{dr}{\cal A} &=&
   \displaystyle
   -K^{-1}{\cal N}_\ell
   V\left[{\cal J}_\ell{\cal A}-{\cal N}_\ell{\cal B}\right]\\[5mm]
   \displaystyle
   \frac{d}{dr}{\cal B} &=&
   \displaystyle
   -K^{-1}{\cal J}_\ell
   V\left[{\cal J}_\ell{\cal A}-{\cal N}_\ell{\cal B}\right]
   \end{array}   
   \right.
\end{equation}
with the boundary conditions that follow from (\ref{regularBC})
\begin{equation}
\label{ABregularBC}
   \lim_{r\to0}{\cal A}_{nn'}(E,r)=2\delta_{nn'}\ ,
   \qquad
   \lim_{r\to0}{\cal B}_{nn'}(E,r)=0\ .
\end{equation}
The representation of $\phi(E,r)$ in terms of ${\cal A}$, ${\cal B}$
and ${\cal F}^{(\rm in/out)}$ is equivalent. However, from a practical
point of view, near the origin it is more convenient to use
Eqs. (\ref{ABMatrixEq}) which do not have the problem of singularity
cancellation and thus are numerically stable. At large distances, we
have to solve Eqs. (\ref{FinFoutMatrixEq}) in order to obtain the Jost
matrices (${\cal A}$ and ${\cal B}$ involving both ${\cal F}^{(\rm
in/out)}$, generally do not converge with any rotation angle).  Thus,
we can follow the same path shown in Fig. \ref{fig.IntegrationPath},
making transformation from $\left\{{\cal A}, {\cal B}\right\}$ to
$\left\{{\cal F}^{(\rm in)},{\cal F}^{(\rm out)}\right\}$ at $r=b$.

In principle, from the very beginning when solving either the
Schr\"odinger equation or Eqs. (\ref{ABMatrixEq}), we could move along
the ray (\ref{rot}). This however could be a source of numerical
instability of another kind. If the potential has exponential
functions, it is oscillating with complex $r$. Moving along the real
axis allows us to avoid this problem. Moreover, the potential may be
given in an analytical form only for large $r$ while in the inner
region it is known in the form of a table along the real axis.

\section{Numerical example}
In order to demonstrate the efficiency and accuracy of the proposed
method, we apply it here to a particular two-channel problem.
Another purpose of the calculations reported further down,  is to
learn how individual resonances contribute to observable scattering
cross sections.

\subsection{Potential}
As a testing ground, we chose the model proposed by Noro And Taylor
\cite{NoroTaylor} and used in many other publications since then.
Their two-channel potential 
\begin{equation}
\label{NTpotential}
    U(r)=\left(
    \begin{array}{cc}
      -1.0 & \ -7.5\\[3mm]
      -7.5 & \ \phantom{+} 7.5
      \end{array}
      \right)r^2e^{-r}\ ,
\end{equation}
is of a short range and apparently obeys the condition
(\ref{Vcondition}).  The units in this model are chosen in such a way
that $\mu_1=\mu_2=\hbar c=1$. In the tables and figures given in the
subsequent Sections, the units for the energies and cross sections are
therefore not indicated.  

The threshold energies are $E_1=0$ and $E_2=0.1$. The potential
(\ref{NTpotential}) has an attractive well in the lower channel, a
barrier in the upper channel, and rather strong coupling between the
two channels. This means that it generates a rich spectrum and
represents a real challenge to any method that is designed to analyze
the roles played by individual resonances in the whole picture of the
scattering process.

All the calculations presented here, were done for $\ell=0$. This is
not because our method has some limitations or needs simplifications.
It can handle any value of the angular momentum (even complex).  Since
however this is an artificial model and there are no partial waves
that would present a special physical interest, we chose the lowest
one. Moreover, to the best of our knowledge, in all other publications
only the $S$-wave scattering was considered and we therefore have to
do the same to be able to compare some results.

\subsection{Spectral points and cross sections}
Using the Runge-Kutta-Fehlberg method \cite{RungeKutta}, we solved
Eqs. (\ref{ABMatrixEq}) from $r_{\rm min}=0.0001$ to $b=1$ along the
real axis. Then with the same numerical method, we solved
Eqs. (\ref{FinFoutMatrixEq}) from $r=b$ to $|r|=30$ along the path
shown in Fig. \ref{fig.IntegrationPath}. For searching the bound
states and calculating the $S$-matrix for real energies, the rotation
angle $\theta$ was taken to be zero. When the resonances were located
and the $S$-matrix was calculated in the complex $E$-plane,
we used appropriate values for this angle (up to $0.4\pi$) that 
turned the unitary cuts (see. Fig. \ref{fig.spectralpoints}) far enough 
to open necessary domains of the complex plane.

\begin{figure}[htp]
\centerline{\epsfig{file=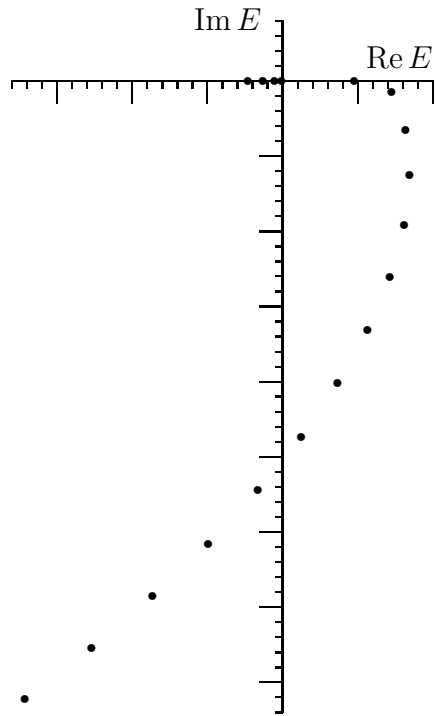}}
\caption{\sf
Spectral points generated by the potential (\ref{NTpotential}) and
given in  Table~\protect\ref{table.spectrum}.   
}
\label{fig.spectrum}
\end{figure}

The spectral points nearest to the origin of the $E$-plane are given
in Table \ref{table.spectrum} where the corresponding results from
some other publications are presented for the sake of comparison.  The
$S$-matrix poles corresponding to these spectral points are shown in
Fig. \ref{fig.spectrum}.  
The potential (\ref{NTpotential}) supports
four bound states and generates a string of nine resonances at energies
above the second threshold. 
All other spectral points are sub-threshold resonances and lie at 
negative energies.

The two lowest bound states lie below the minimum of the attractive 
potential in the first channel.
This shows that the system under consideration is essentially diabatic
as one should expect with strong cross-channel coupling in the
potential (\ref{NTpotential}).

The $S$-wave scattering cross sections for the transitions from
channel $n$ to channel $n'$,
\begin{equation}
\label{crossection}
      \sigma(n\to n')=\frac{\pi}{k_n}|\delta_{n'n}-S_{n'n}|^2\ ,
\end{equation}
presented in Figs. \ref{fig.sigma11}, \ref{fig.sigma12}, and
\ref{fig.sigma22}, show several bumps and peaks.  Our traditional
understanding of quantum scattering, preoccupied with the Breit-Wigner
picture, would push us towards associating these cross section
irregularities with individual resonance poles of the $S$-matrix. Let
us not jump to such conclusions too hastily.  Having analyzed how the
cross section is built up out of individual contributions from the
resonances, in the subsequent sections we will see that the
$\sigma(n\to n')$ curves are the results of rather complicated
interplay of all the poles and therefore a Breit-Wigner
parametrization of them would be naive and misleading.

\subsection{Mittag-Leffler expansion}
The Mittag-Leffler theorem \cite{MittagLeffler} offers a way to
decompose the $S$-matrix in a sum of terms representing its poles and
the background. The theorem itself is rather general. In plain words,
it says that a meromorphic function can be expanded in a sum of an
entire function and a series of its principal parts at all the poles.
What we actually need is more simple.  Leaving the mathematical
subtleties,
we want to expand the $S$-matrix in a
sum of few singular terms (poles) and something that takes into
account everything else (background term).  This can be done using the
Cauchy integral formula. Indeed, consider a contour shown in
Fig. \ref{fig.contour}, which encloses $N$ poles of the $S$-matrix.
Then the Cauchy formula reads
\begin{equation}
\label{CauchyIntegral}
    \oint \frac{S(\zeta)}{\zeta-E}d\zeta=
    2\pi i S(E)+2\pi i\sum_{j=1}^N
    \frac{{\rm Res}[S(E_j)]}{E_j-E}\ ,
\end{equation}
where $E$ is a point inside the contour and $E_j$ are the resonance
energies. Therefore in the desired expansion,
\begin{equation}
\label{MLexpansion}
    S(E)= \sum_{j=1}^N
    \frac{{\rm Res}[S(E_j)]}{E-E_j}+  
    \frac{1}{2\pi i}\oint \frac{S(\zeta)}{\zeta-E}d\zeta\ ,
\end{equation}
the $N$ selected poles are represented explicitly while the rest of
them as well as the background (the Mittag-Leffler's entire function)
are taken into account via the contour integral.  In order to
understand what role a particular resonance plays in the scattering
process, we will simply omit the corresponding term from the sum
(\ref{MLexpansion}) and analyze the consequences.
In a sense, the expansion (\ref{MLexpansion}) is similar to that of 
Ref.\cite{Brandas} where the spectral-density function $m^+(\lambda)$
was presented as a sum over the resonances and the background term.
It should be noted, however, that Eq. (\ref{MLexpansion}) is not an
approximation but an exact expansion when all $N$ poles enclosed by
the contour are taken into account.

To calculate the $S$-matrix via Eq. (\ref{MLexpansion}), we need
to know its residues at the poles and to be able to calculate the
integral for a given energy $E$. The residues can be found by numerical
differentiation of the Jost matrix determinant. Indeed, according to
Eq. (\ref{ASmatrdef})
\begin{equation}
\label{SmatrixExplicit}
    S=f^{(\rm out)}\left(
    \begin{array}{cc}
    f^{(\rm in)}_{22} & -f^{(\rm in)}_{12} \\[3mm]
    -f^{(\rm in)}_{21} & f^{(\rm in)}_{11}
    \end{array}\right)
    \frac{1}{\det f^{(\rm in)}}\ ,
\end{equation}
where the poles are caused by simple zeros of $\det f^{(\rm in)}(E)$.
Therefore
\begin{equation}
\label{ResidueExplicit}
    {\rm Res}\,[S(E)]=f^{(\rm out)}(E)\left(
    \begin{array}{cc}
    f^{(\rm in)}_{22}(E) & -f^{(\rm in)}_{12}(E) \\[3mm]
    -f^{(\rm in)}_{21}(E) & f^{(\rm in)}_{11}(E)
    \end{array}\right)
    \left[\frac{d}{dE}\det f^{(\rm in)}(E)\right]^{-1}
\end{equation}
with the derivative numerically obtained within the central difference
approximation
\begin{equation}
\label{DerivativeExplicit}
   \frac{d}{dE}\det f^{(\rm in)}(E)\approx
   \frac{\det f^{(\rm in)}(E+\epsilon)-\det f^{(\rm in)}(E-\epsilon)}
   {2\epsilon}\ .
\end{equation}
In the calculations, we used $\epsilon=10^{-6}$ which gave the
accuracy of at least 5 digits. This was sufficient for the purpose of
visual comparing the curves when analyzing the contribution to the
total cross section from different $S$-matrix poles.

Thus calculated residues of the $S$-matrix are given in Table
\ref{table.residues} and plotted in Fig. \ref{fig.residues}.
It is interesting to notice that they follow regular pattern in the
form of anti-clockwise spiral. The reason for growth of the spiral
radius is that according to Eq. (\ref{ASnearEres})
$$
     {\rm Res}\,[S_{nn'}(E)]\ \sim\ \sqrt{\Gamma_n\Gamma_{n'}}
$$ 
and the widths become larger every time we go to the next
resonance. The reason for the anti-clockwise turning of the
corresponding phase factor, however, remains unclear. As is
seen from Fig.  \ref{fig.res_phas}, the phases of the residues show
the tendency to converge to a constant angle, which means that for the
far away resonances the points on the spiral become closer and closer
to each other.

The contour integral of Eq. (\ref{MLexpansion}), enclosing nine
($N=9$) resonance points (see Fig.  \ref{fig.contour}), was calculated
using 175 points of the Gaussian quadrature formula on each of the
three line segments that form the contour. With this number of
quadrature points we achieved four digit accuracy which was more than
sufficient for our purpose of visual comparing the cross section
curves. The vertices of the triangle contour (0.5, 0.5), (0.5,-30.0),
and (20.1, 0.5) were chosen in such a way that both branching points
are outside of it and with $\theta\ge 0.25\pi$ the contour does not
cross the unitary cuts.

A question may arise: Why did we not include the bound state poles in
the contour? There are two reasons for this. Firstly, the present work
was motivated (as we explained in the Introduction) by the controversy
in associating the irregularities of the chemical reaction cross
sections with intermediate resonant states. Therefore, our goal was to
develop a technique for examining the contributions of the resonance
poles, not the bound states.  Secondly, a contour enclosing the bound
states, would require calculating the $S$-matrix via
Eq. (\ref{ASmatrdef}) at the points around them, i.e. with ${\rm
Im\,}k_n>0$. At such points, however, the limit (\ref{fOutDefinition})
does not exist. In order to reach these points, we would need a
negative rotation angle with $\theta<-\pi/2$ which is not possible.
There is a way to circumvent this difficulty by expanding the
solutions ${\cal F}^{(\rm in/out)}$ in series near $E=0$ (see
Ref. \cite{rakpup}), but this would require a separate publication. It
should be emphasized that the inclusion of only nine poles in the
contour is not an approximation. The Cauchy theorem
(\ref{CauchyIntegral}) is exact, provided that all enclosed poles are
taken into account. All the other poles are taken care of by the
contour integral.

\subsection{Analysis of the cross sections}
Before starting the analysis, we tested the numerical procedure by
comparing the cross sections directly obtained from
Eqs. (\ref{FinFoutMatrixEq}) at real energies with $\theta=0$ and the
corresponding cross sections calculated from the expansion
(\ref{MLexpansion}). The same $\sigma(n\to n')$ curves, namely, shown
in Figs. \ref{fig.sigma11}, \ref{fig.sigma12}, and \ref{fig.sigma22},
were obtained in both cases (within four digit accuracy, which is
visually indistinguishable).  Having thus established that the sum
(\ref{MLexpansion}) reproduces the cross sections correctly, we
started to investigate the importance of its terms by excluding them
from the sum and calculating the resulting cross sections.

First of all, we omitted all the nine poles. To our surprise, the
remaining integral part of the expansion (\ref{MLexpansion}) did not
turn out to be a smooth function. Contrary to our expectation, at low
energies it gives almost the same structure of peaks and deeps as they
are in the exact cross sections (see Figs. \ref{fig.0cs11},
\ref{fig.0cs12}, and \ref{fig.0cs22}). This means that these peaks are
of a non-resonance origin. They are either the threshold
irregularities (known in scattering theory as threshold
cusps\cite{baz,Landau}), or the influence of some other $S$-matrix
poles 
(especially the four bound states that are close to the point
$E=0$).

In an attempt to understand what role each individual resonance plays
in the scattering process, we excluded them one at a time from the sum
(\ref{MLexpansion}). This procedure can give us some indication and
hints.  In Figs. \ref{fig.out.11}, \ref{fig.out.12}, and
\ref{fig.out.22} the cross sections $\sigma(1\to1)$, $\sigma(1\to2)$,
and $\sigma(2\to2)$ thus obtained are shown for eight different
choices of the resonance to be excluded. 

These figures do not show the effect of the exclusion of the first
resonance because this effect is very simple. When the first resonance
is omitted, the narrow deep in $\sigma(1\to2)$ and peak in
$\sigma(2\to2)$ at $E=4.768$ disappear while the other parts of the
cross section curves remain unchanged. As far as the elastic cross
section $\sigma(1\to1)$ is concerned, the resonant cusp at $E=4.768$
(see insert on Fig. \ref{fig.sigma11}) is so weak that it is hardly
visible and therefore the exclusion of the first resonance pole from
the sum (\ref{MLexpansion}) is practically unnoticeable. The reason
for that can be found in Table \ref{table.residues}. The $S_{11}$
element of the $S$-matrix has extremely small residue at this pole. As
a result the corresponding term in the sum (\ref{MLexpansion}) for
real $E$ is always negligible. Being used to the Breit-Wigner picture,
we expected to have a sharp energy dependence at around $E\sim 4.8$ in
all channels. To our surprise, this turned out to be not the case
because of smallness of ${\rm Res}(S_{11})$.

Apart from the first resonance ($E=4.768$), 
the only place where we can see a clear manifestation of a resonance
through a distinct peak is the top left corner of
Fig. \ref{fig.out.12}, where one of the peaks in the inelastic cross
section $\sigma(1\to2)$ disappears together with the second resonance
pole while the rest of the curve retains its shape.  As far as the
elastic scattering in the channels 1 and 2 is concerned (see Figs.
\ref{fig.out.11} and \ref{fig.out.22}), the only information we can
deduce from the curves is that the second and the third resonances are
responsible for the bumps between $E=5$ and $E=10$, the far resonances
do scaling of the curves, and no resonances manifest themselves
through distinct peaks 
(except for the first one, of course). 
This is from a general physical view understandable in the sense that
the fourth and higher resonances are so wide that they, when excited,
affect a rather wide energy region.

In general, the cross section analysis based on the expansion
(\ref{MLexpansion}), is not straight forward. This is not surprising
because the contributions of the terms of this expansion into the
cross section are not linear. Indeed, on the right hand side of
Eq. (\ref{crossection}) in addition to squares of these terms there
are many interference products whose contributions depend on relative
phases of the terms. Therefore the full cross section is a result of
complicated interplay of all resonances and the background term.  The
analysis of the sum (\ref{MLexpansion}) itself, i.e. examining the
contributions of the resonances into the $S$-matrix, could shed some
light on the role of individual resonances in the scattering picture.
The most convenient way of doing this is the analysis of the Argand
plots which is done next.

\subsection{Argand plots}
The Argand plots are widely used to depict complex valued functions
depending on a real parameter. For each value of this parameter, the
function value corresponds to a point in the complex plane. If the
function is continuous, the set of such points forms a curve.

In the scattering theory, the term Argand plot has a specific meaning,
namely, the curve in the complex plane, along which the $S$-matrix
element moves when the (real) collision energy
increases. Traditionally the Argand plots were mainly used in particle
physics. There are however many examples of their application in
atomic and molecular physics as well \cite{kuppermann,aoiz}.

It can be shown (see, for example, Refs. \cite{kuppermann} and
\cite{dalitz}) that in the absence of resonances the phases
(arguments) of elements of the $S$-matrix monotonously decrease with
the increasing energy of collision. Therefore these matrix elements as
functions of real energy move along circular trajectories in the
clockwise direction on the Argand plot. Contrary to that, a resonance
causes the scattering phase shift to increase by $\pi$, which means an
increase of the $S$-matrix phase by $2\pi$ (i.e. full anti-clockwise
circle). This is, of course, an idealized picture for an isolated
resonance with weak influence of the background scattering.

In a vicinity of a resonance, the $S$-matrix can always be split in
two terms: The background and resonance terms, with decreasing and
increasing phases, respectively. Their interplay may cause the
curvature of the Argand trajectory to change.  If a resonance is
strong, the curvature changes so much as making the point on the curve
to move along an arc in the anti-clockwise direction.  With a weak
resonance, the curve changes its curvature only slightly.

\begin{figure}[htp]
\centerline{\epsfig{file=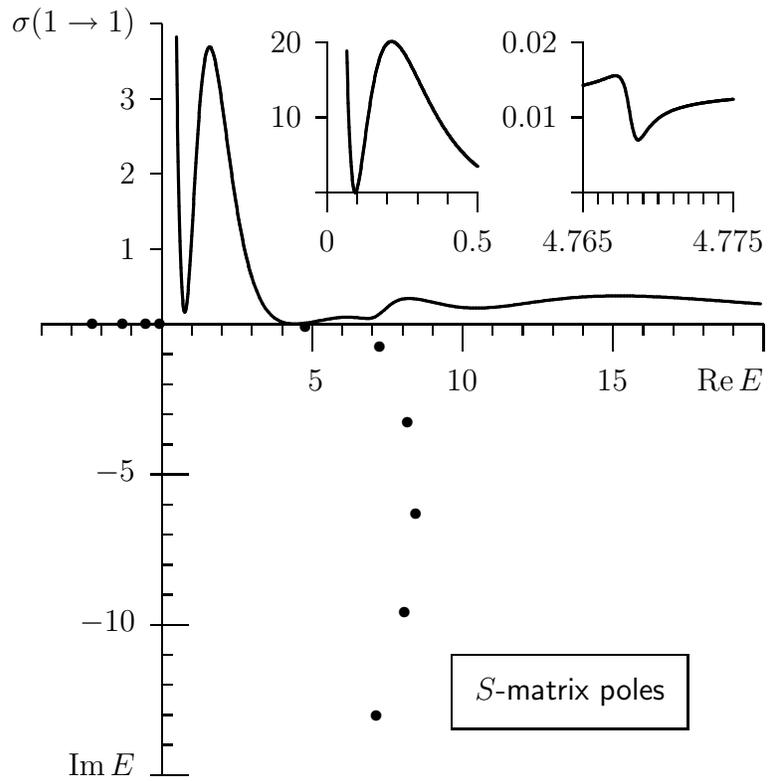}}
\caption{\sf
Energy dependence of the elastic scattering cross section in channel 1
for the potential (\ref{NTpotential}). Few of the $S$-matrix poles
(see Table~\protect\ref{table.spectrum} and
Fig.~\protect\ref{fig.spectrum}) are shown in the lower part of the
Figure.
}
\label{fig.sigma11}
\end{figure}

\begin{figure}[htp]
\centerline{\epsfig{file=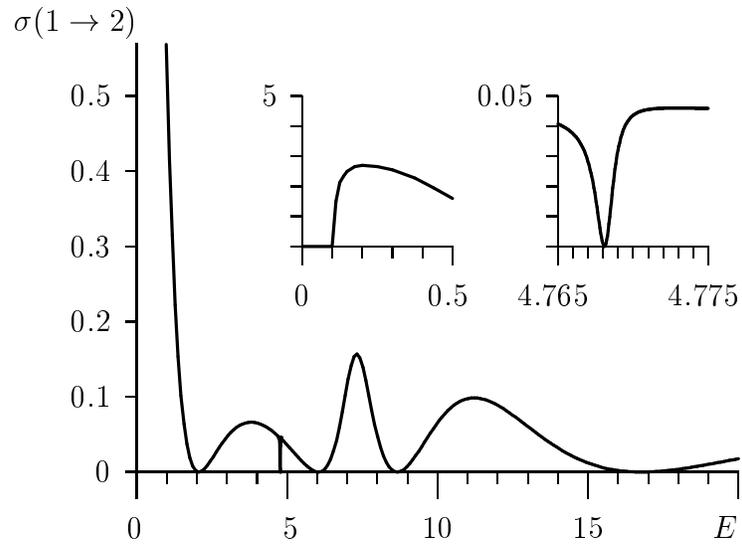}}
\caption{\sf
Cross section energy dependence of the inelastic transition 
($1\to2$) for the potential (\ref{NTpotential}). 
}
\label{fig.sigma12}
\end{figure}

\begin{figure}[htp]
\centerline{\epsfig{file=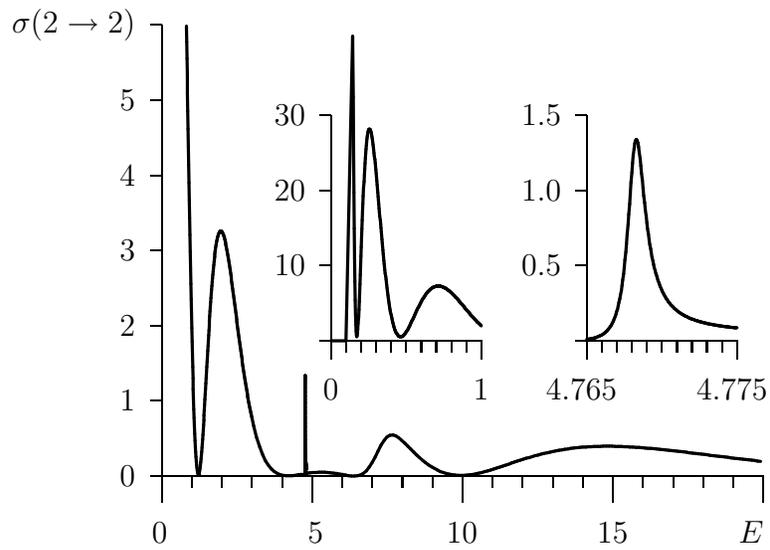}}
\caption{\sf
Energy dependence of the elastic scattering cross section in channel 2
for the potential (\ref{NTpotential}).
}
\label{fig.sigma22}
\end{figure}

\begin{figure}[htp]
\centerline{\epsfig{file=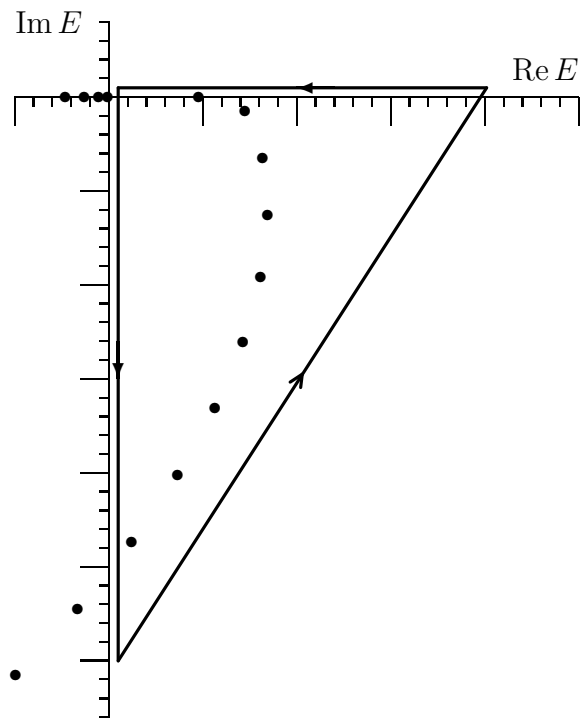}}
\caption{\sf
The integration contour for Eq. (\ref{CauchyIntegral}). The vertices of
the triangle are (0.5,0.5), (0.5,$-30$), and (20.1,0.5).
}
\label{fig.contour}
\end{figure}

\begin{figure}[htp]
\centerline{\epsfig{file=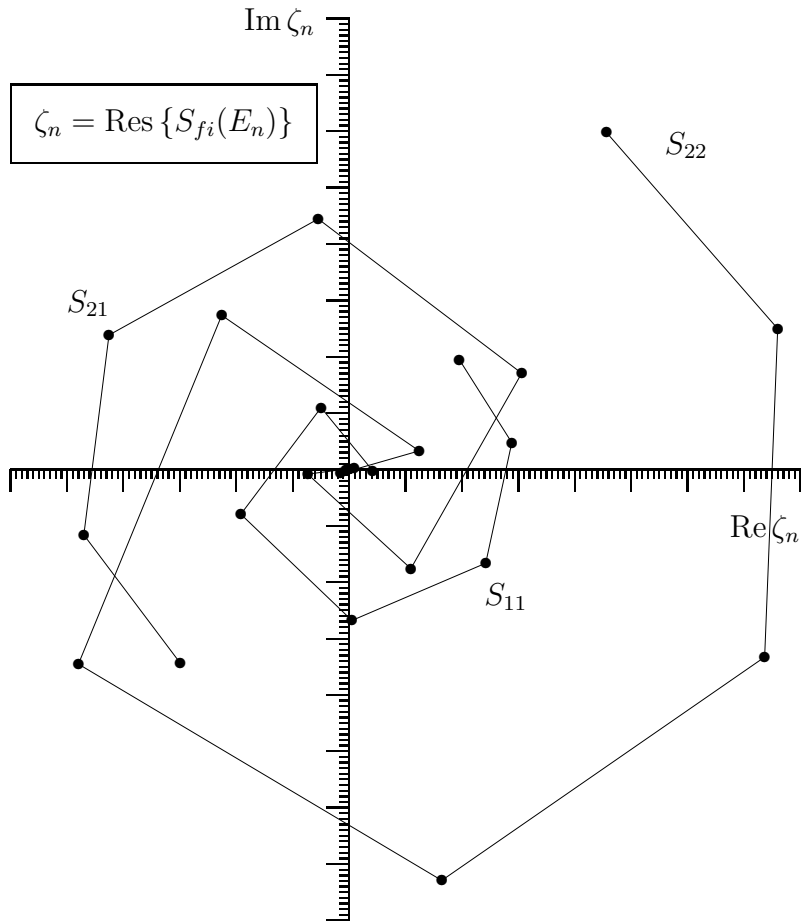}}
\caption{\sf
Residues of the $S$-matrix elements at the 9 resonance poles enclosed
in the contour shown in Fig.~\protect\ref{fig.contour}. The
coordinates of the points are given in Table~\protect\ref{table.residues}. 
}
\label{fig.residues}
\end{figure}

\begin{figure}[htp]
\centerline{\epsfig{file=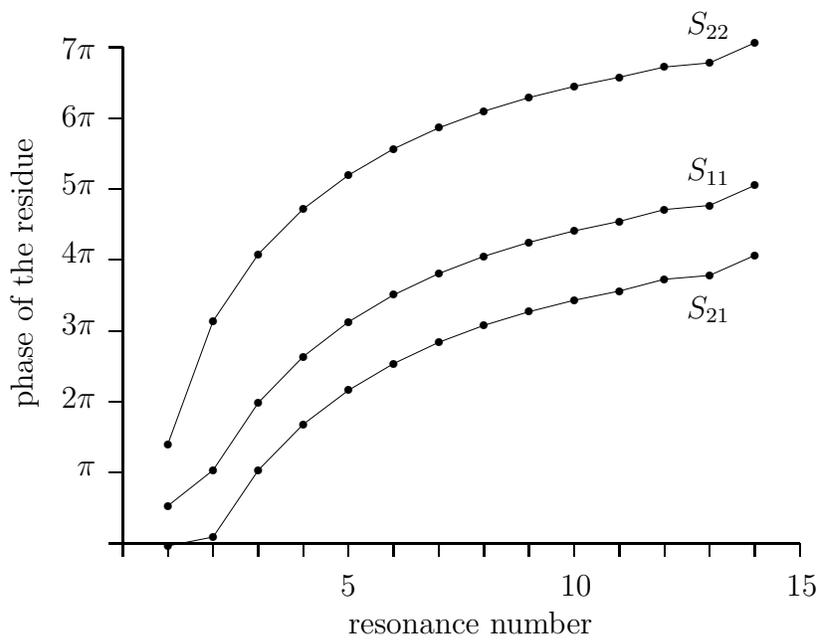}}
\caption{\sf
Phases of the residues of the $S$-matrix elements at the first 14
resonance poles given in Table~\protect\ref{table.spectrum}. 
}
\label{fig.res_phas}
\end{figure}

\begin{figure}[htp]
\centerline{\epsfig{file=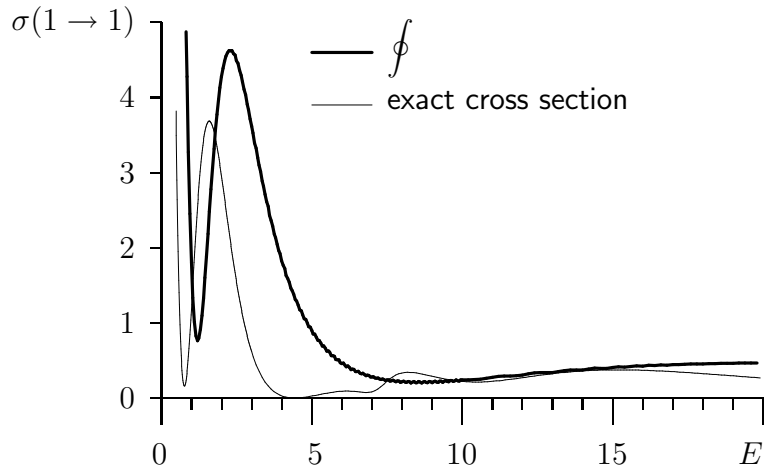}}
\caption{\sf
Comparison of the exact cross section for the elastic scattering in
the channel 1 (thin curve) with the corresponding cross section 
obtained from the
expansion (\ref{MLexpansion}) where all the pole terms are omitted 
(thick curve).
}
\label{fig.0cs11}
\end{figure}

\begin{figure}[htp]
\centerline{\epsfig{file=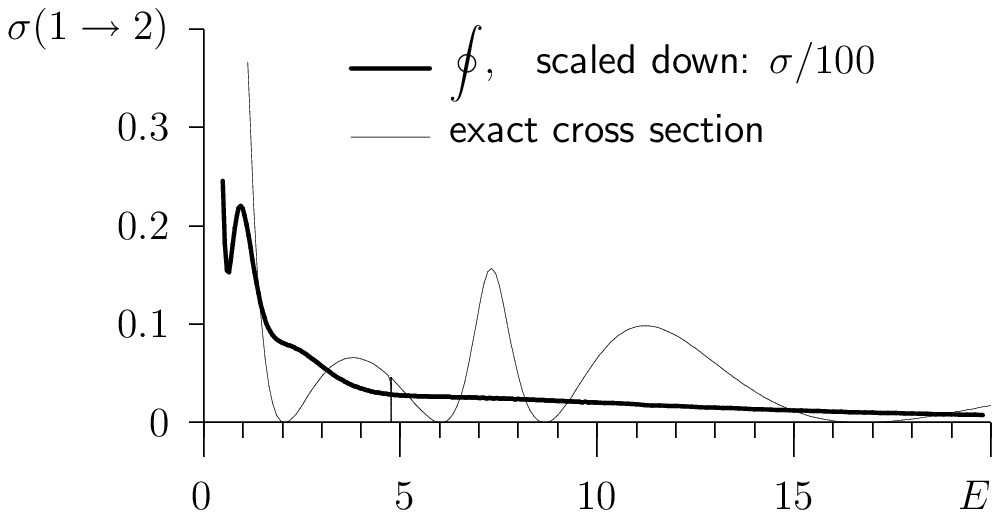}}
\caption{\sf
Comparison of the exact cross section for the inelastic transition $(1\to2)$
(thin curve) with the corresponding cross section obtained from the
expansion (\ref{MLexpansion}) where all the pole terms are omitted. 
In order to fit into the picture the thick curve is scaled down by the
factor of 1/100.
}
\label{fig.0cs12}
\end{figure}

\begin{figure}[htp]
\centerline{\epsfig{file=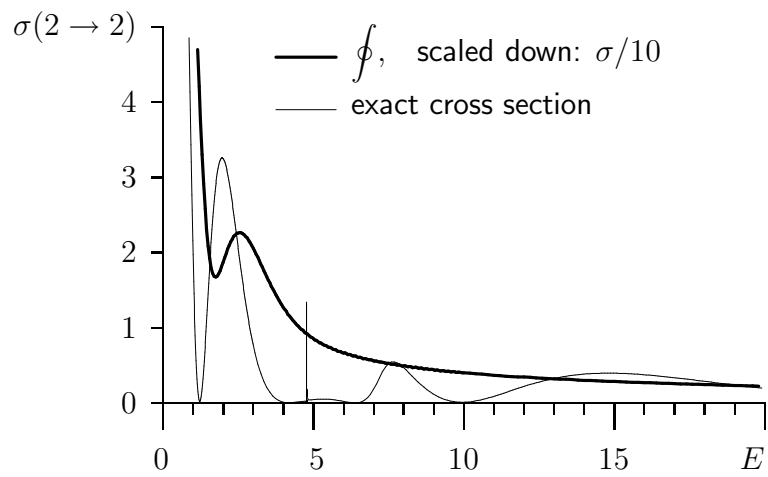}}
\caption{\sf
Comparison of the exact cross section for the elastic scattering in
the channel 2 (thin curve) with the corresponding cross section
obtained from the expansion (\ref{MLexpansion}) where all the pole
terms are omitted. In order to fit into the picture the thick curve is
scaled down by the factor of 1/10.
}
\label{fig.0cs22}
\end{figure}

\begin{figure}[htp]
\centerline{\epsfig{file=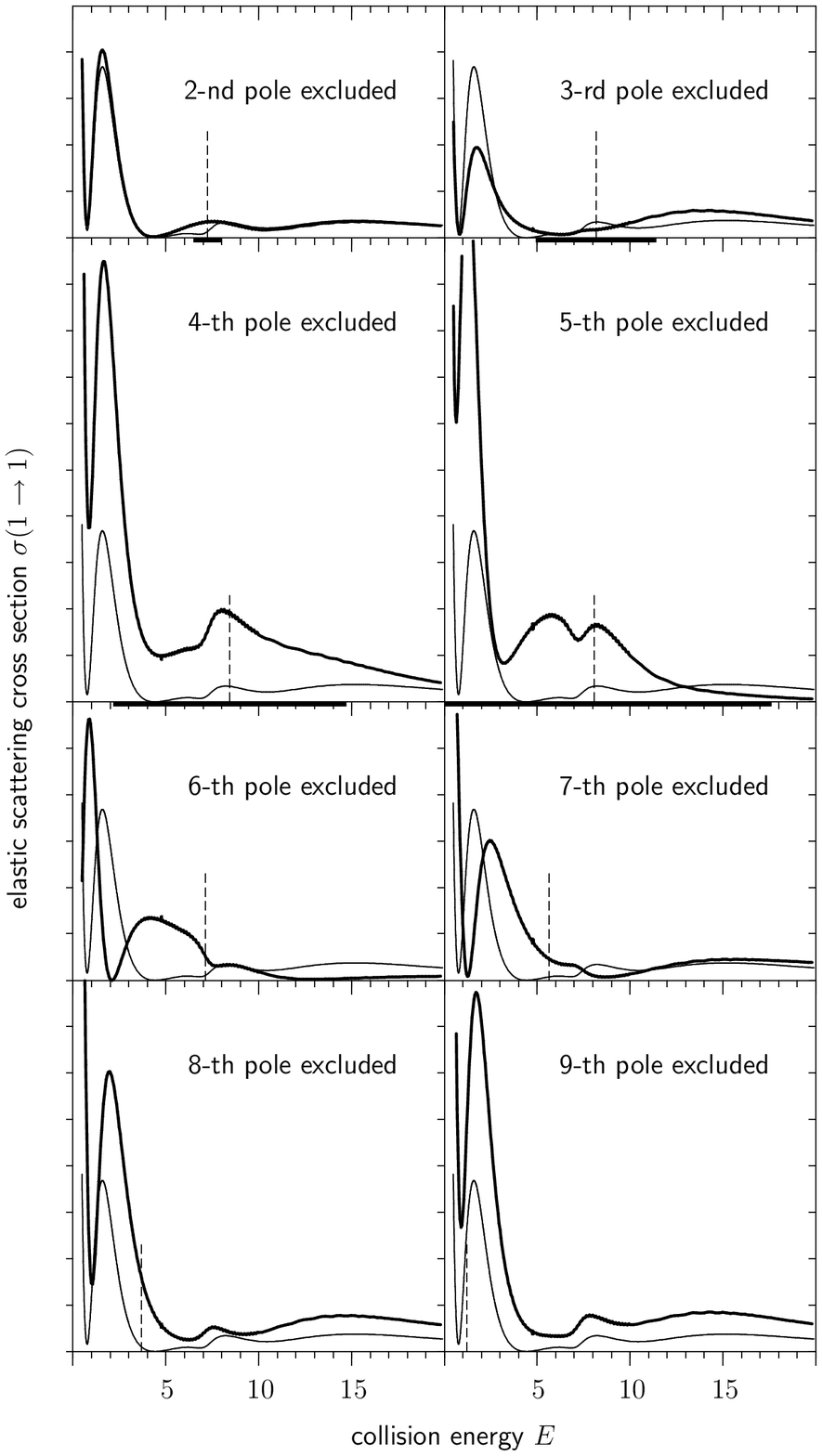}}
\caption{\sf
Exact cross section for the elastic scattering in channel 1 (thin
curve) and the corresponding cross section obtained from the expansion
(\ref{MLexpansion}) where one of the nine pole terms is omitted 
(thick curve). Vertical dashed lines indicate the positions $E_{\rm r}$ 
of the excluded resonances and thick horizontal bars on the energy axis 
cover the corresponding intervals $E_{\rm r}\pm \Gamma/2$. For the last 
four resonances, such intervals exceed the energy segment shown on the
figure.
}
\label{fig.out.11}
\end{figure}

\begin{figure}[htp]
\centerline{\epsfig{file=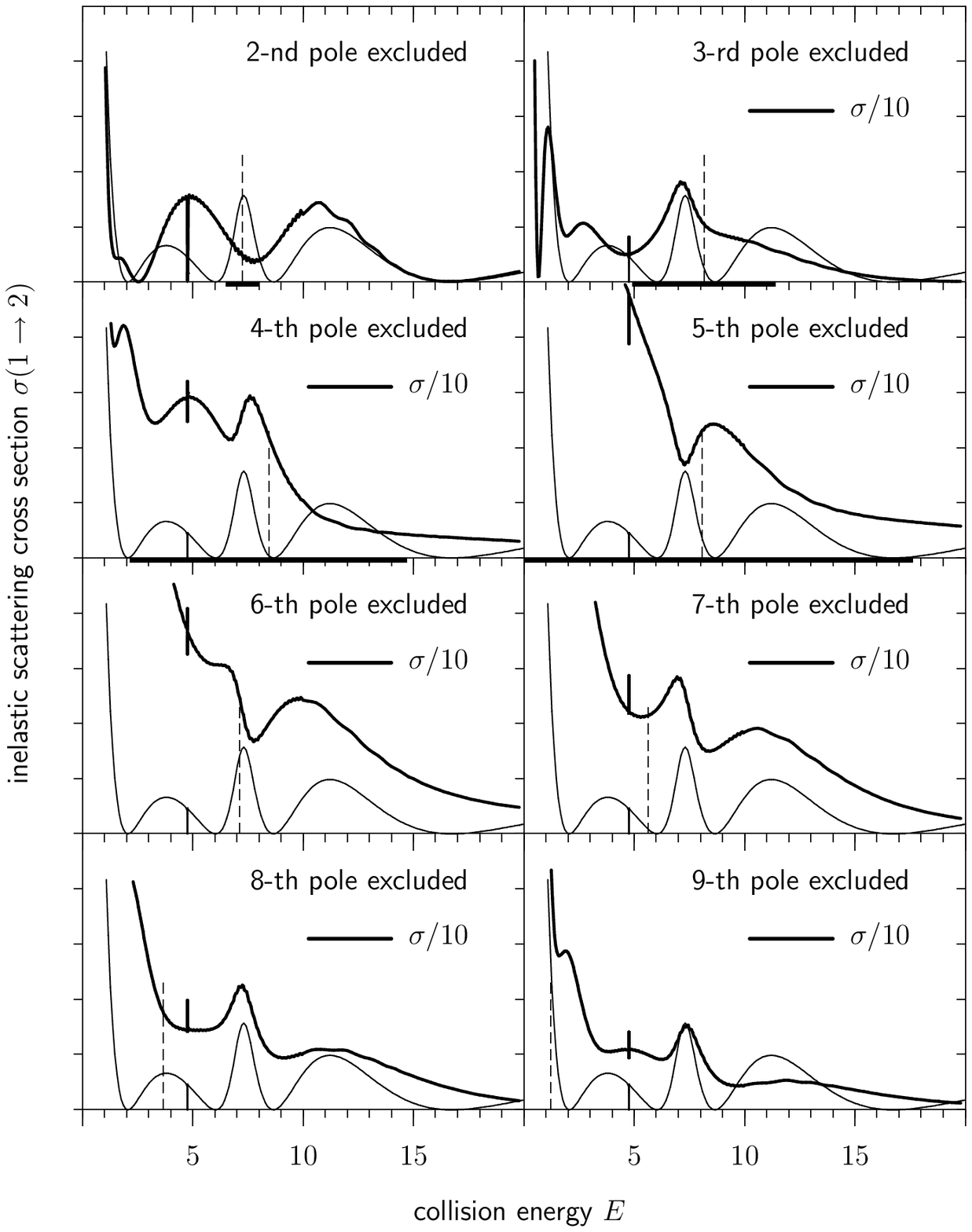}}
\caption{\sf
Exact cross section for the inelastic transition $1\to2$ (thin
curve) and the corresponding cross section obtained from the expansion
(\ref{MLexpansion}) where one of the nine pole terms is omitted 
(thick curve). As indicated on the graphs, some of the thick curves are
scaled down by the factor of $0.1$ in order to fit into the picture.
Vertical dashed lines indicate the positions $E_{\rm r}$ 
of the excluded resonances and thick horizontal bars on the energy axis 
cover the corresponding intervals $E_{\rm r}\pm \Gamma/2$. For the last 
four resonances, such intervals exceed the energy segment shown on the
figure. 
}
\label{fig.out.12}
\end{figure}

\begin{figure}[htp]
\centerline{\epsfig{file=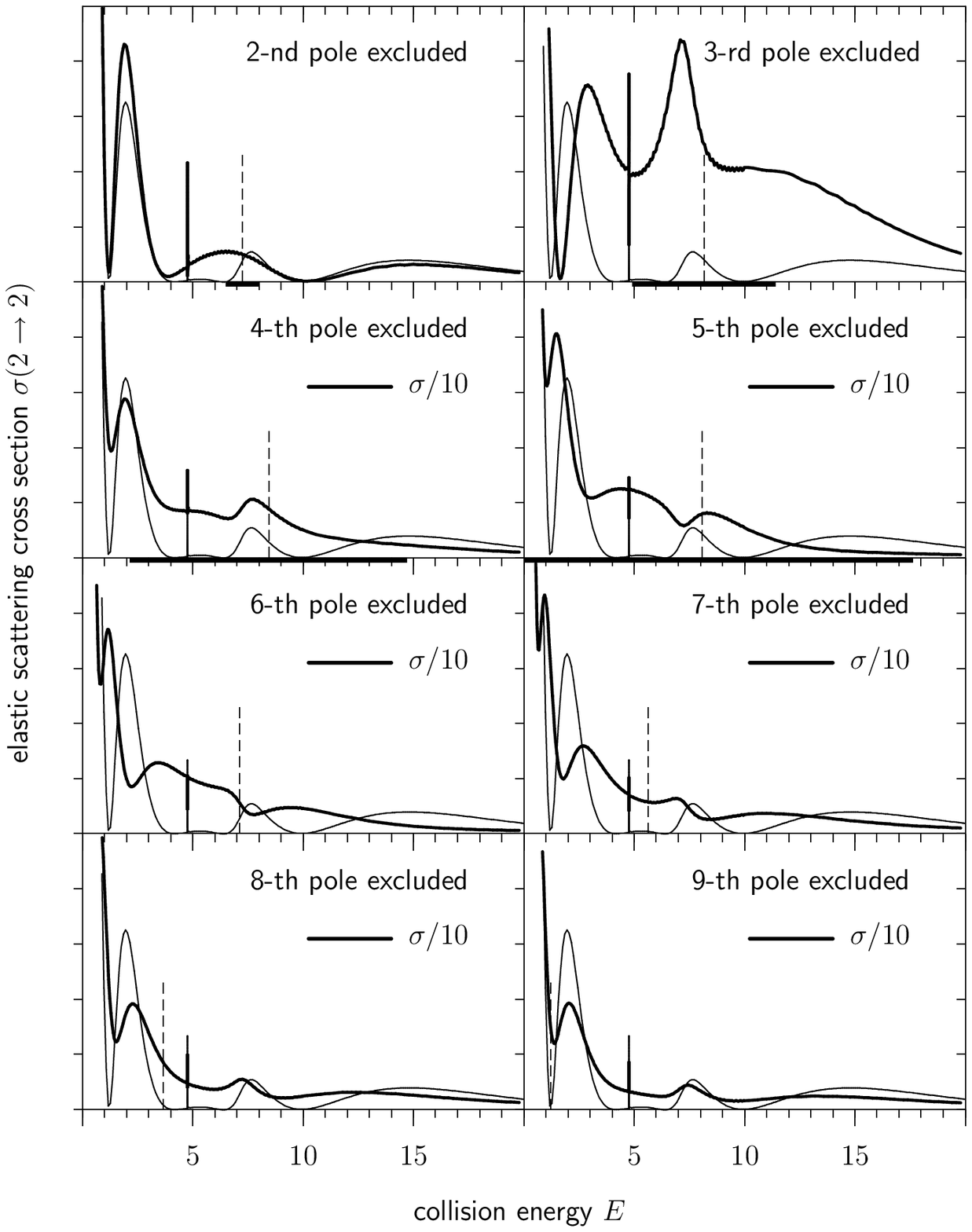}}
\caption{\sf
Exact cross section for the elastic scattering in channel 2 (thin
curve) and the corresponding cross section obtained from the expansion
(\ref{MLexpansion}) where one of the nine pole terms is omitted 
(thick curve). As indicated on the graphs, some of the thick curves are
scaled down by the factor of $0.1$ in order to fit into the picture. 
Vertical dashed lines indicate the positions $E_{\rm r}$ 
of the excluded resonances and thick horizontal bars on the energy axis 
cover the corresponding intervals $E_{\rm r}\pm \Gamma/2$. For the last 
four resonances, such intervals exceed the energy segment shown on the
figure.
}
\label{fig.out.22}
\end{figure}

\begin{figure}[htp]
\centerline{\epsfig{file=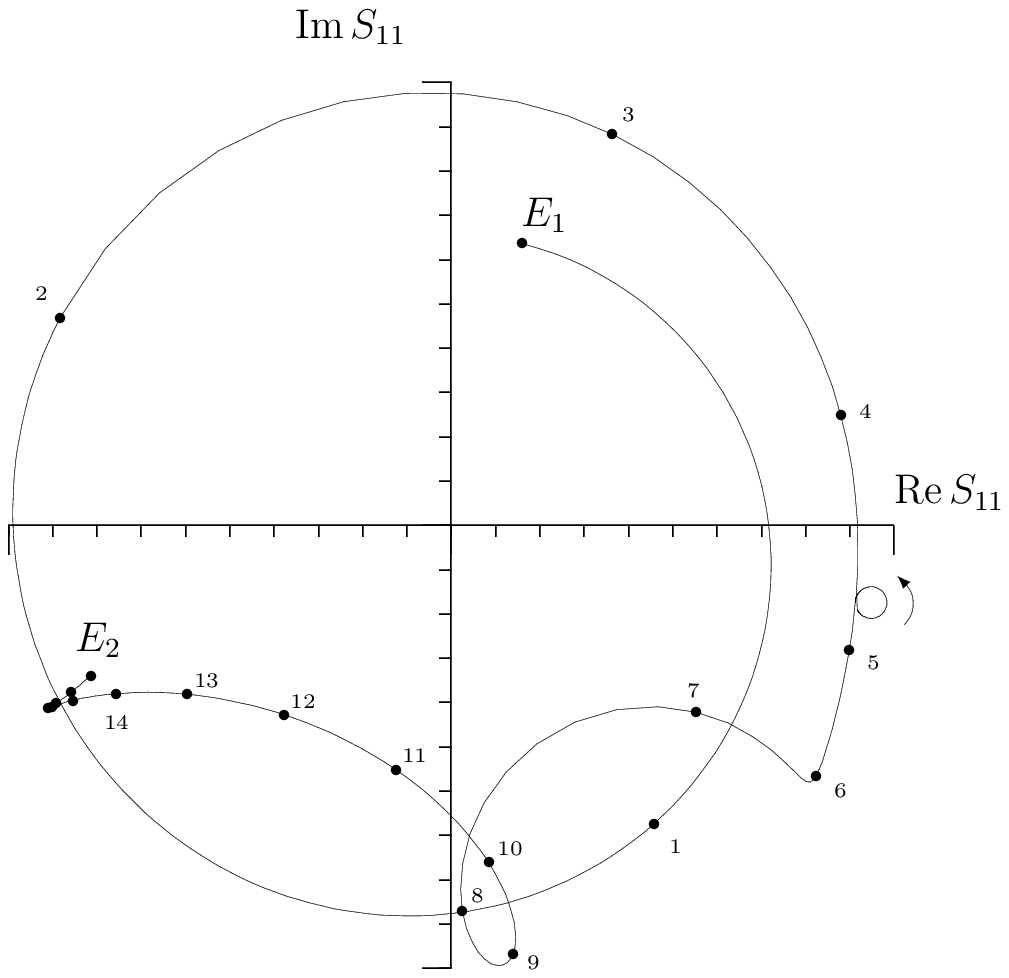}}
\caption{\sf
Argand plot of $S_{11}(E)$ in the energy interval from $E_1=0.5$ to
$E_2=20$. The dots on the curve indicate the points corresponding to
$E=1, 2, 3, \dots$.
}
\label{fig.argand.11}
\end{figure}

\begin{figure}[htp]
\centerline{\epsfig{file=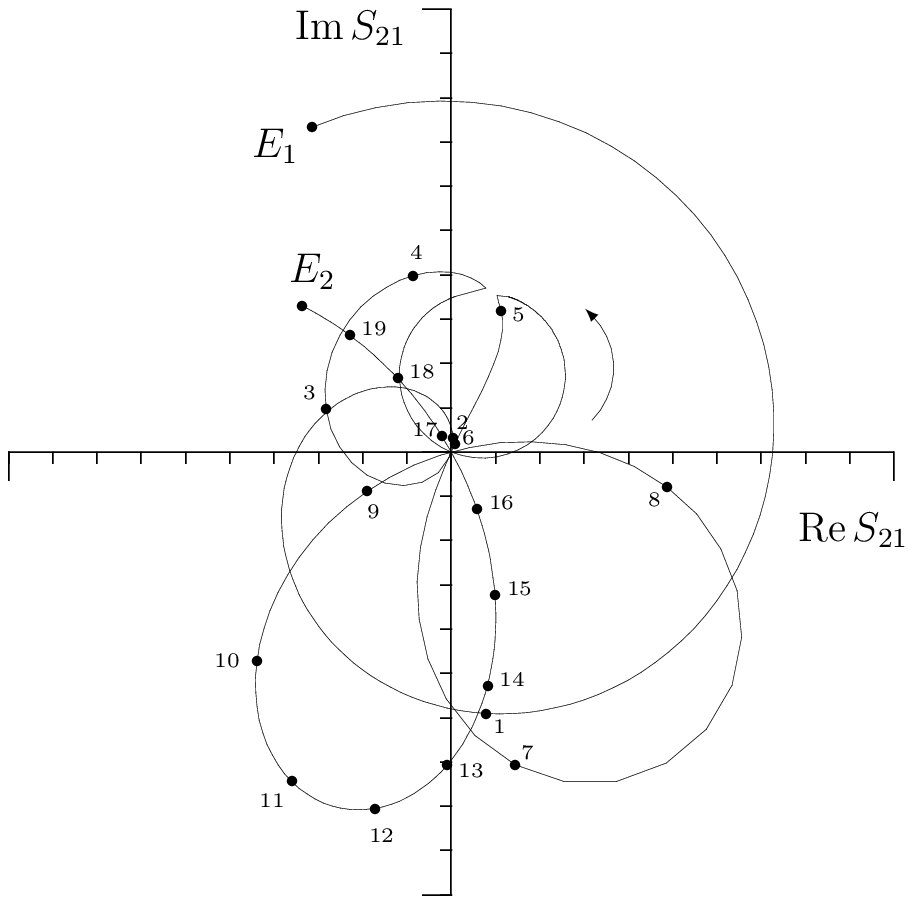}}
\caption{\sf
Argand plot of $S_{21}(E)$ in the energy interval from $E_1=0.5$ to
$E_2=20$. The dots on the curve indicate the points corresponding to
$E=1, 2, 3, \dots$.
}
\label{fig.argand.21}
\end{figure}

\begin{figure}[htp]
\centerline{\epsfig{file=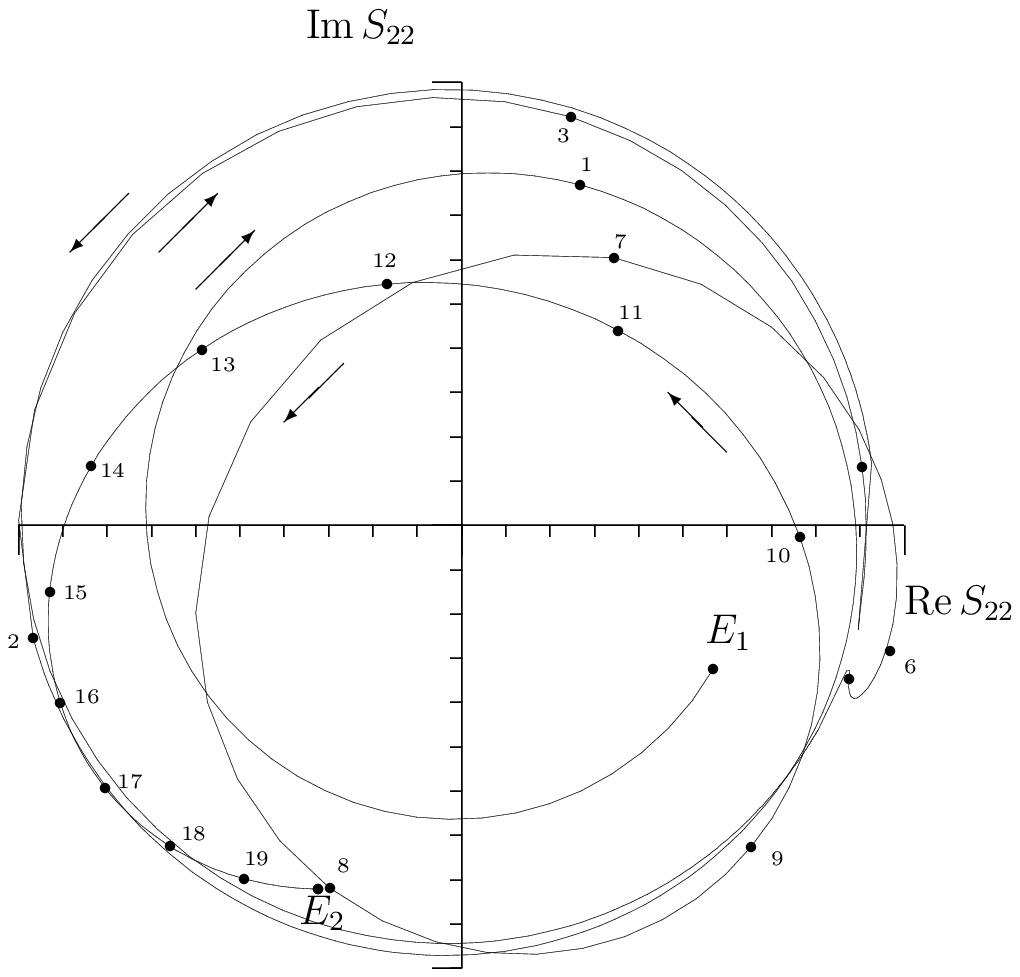}}
\caption{\sf
Argand plot of $S_{22}(E)$ in the energy interval from $E_1=0.5$ to
$E_2=20$. The dots on the curve indicate the points corresponding to
$E=1, 2, 3, \dots$.
}
\label{fig.argand.22}
\end{figure}

In Figs. \ref{fig.argand.11}, \ref{fig.argand.21}, and
\ref{fig.argand.22} the Argand plots for the $S$-matrix generated by
the potential (\ref{NTpotential}) are shown in the energy interval
$0.5\le E\le 20$. The dots on the curves mark the integer values of
the energy, namely, $E=1, 2, 3, \dots$. 
As is seen, all three matrix elements have three anti-clockwise arcs
in the intervals $4.765< E < 4.775$, $6< E < 9$, and $9< E < 20$. This
means that the corresponding peaks at $E=4.768$ and 
bumps seen between $E\sim 6$ and $E\sim 20$ in
Figs. \ref{fig.sigma11}, \ref{fig.sigma12}, and \ref{fig.sigma22}, are
of the resonance origin, despite the fact that not all of them are
high or sharp. The only question remains: Which poles are responsible
for these resonances?

To answer this question, we make use of the expansion (\ref{MLexpansion}).
There is no problem of interference here because, in contrast to the
cross section, the contribution of each pole to the $S$-matrix is
linear. Therefore, excluding them one at a time, we can find out which
poles the two anti-clockwise arcs of the Argand plots are associated
with.

\begin{figure}[htp]
\centerline{\epsfig{file=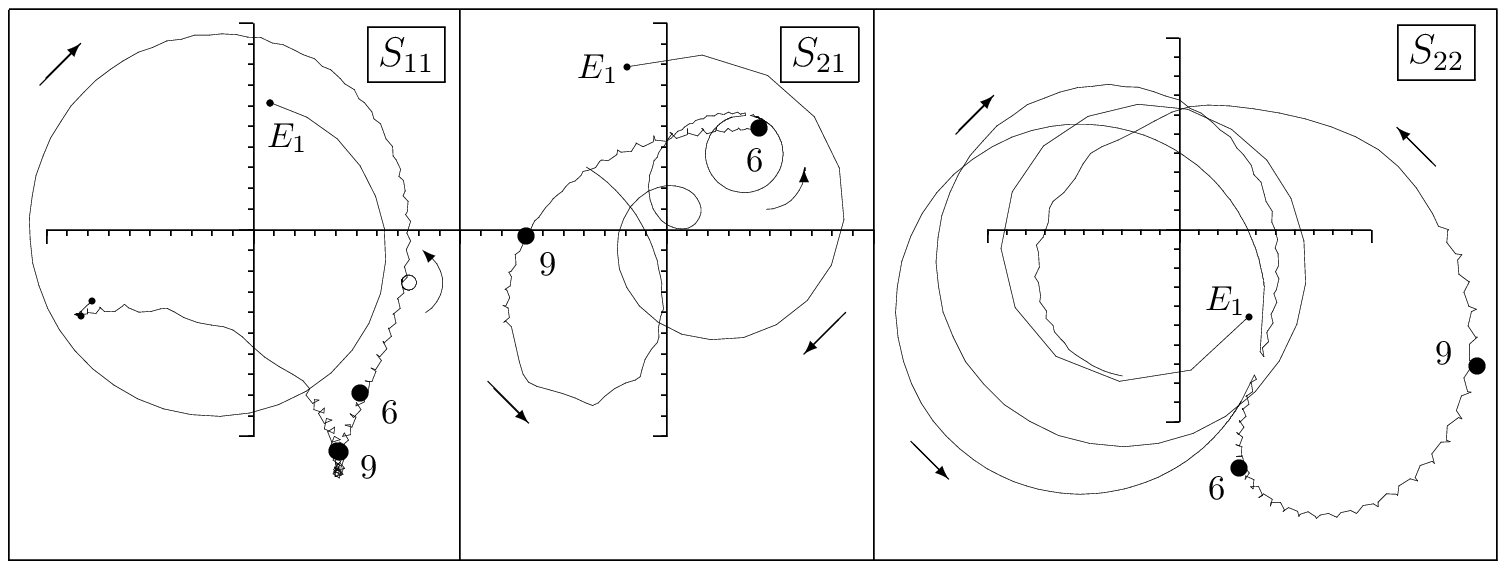}}
\caption{\sf
Argand plot of the $S$-matrix with the second pole excluded. 
The dots mark the points corresponding to  $E=6$ and $E=9$.
}
\label{fig.argand.2out}
\end{figure}

First of all, as was mentioned before, the first resonance pole does
not contribute to anything when $E$ moves away from the point
$E=4.768$.  This is because the denominator of the corresponding term
in the sum (\ref{MLexpansion}) becomes much lager than the residue.
The omission of the first resonance term results therefore in the
disappearance of the first anti-clockwise circles ($4.765< E < 4.775$)
in Figs. \ref{fig.argand.11}, \ref{fig.argand.21}, and
\ref{fig.argand.22}. The rest of these figures remain unaffected.
Contrary to that, the second pole makes a noticeable
contribution everywhere. 
In Fig. \ref{fig.argand.2out} the Argand plots for the
three elements of the $S$-matrix are shown when this pole is excluded
from the sum (\ref{MLexpansion}). It is seen that in all the channels
there is no second anti-clockwise arc between 
$E\sim 6$ and $E\sim 9$ as
compared with Figs.  \ref{fig.argand.11}, \ref{fig.argand.21}, and
\ref{fig.argand.22}. We therefore can conclude that the anti-clockwise
arc in the interval $6< E < 9$ is associated with the second pole and
the bumps in this energy interval on the cross sections shown in Figs.
\ref{fig.sigma11}, \ref{fig.sigma12}, and \ref{fig.sigma22}, are the
manifestation of the second resonance pole, namely, at
$E=7.241200-(i/2)1.511912$.

It is seen that with the exclusion of the second resonance pole, we still
have anti-clockwise arcs for $S_{11}(E)$, starting from $E\sim 9$, and
for $S_{21}(E)$ and $S_{22}(E)$, starting from $E\sim 6$ and extending
to $E\sim 20$. This means that one or several other resonance
poles are responsible for their formation. 

\begin{figure}[htp]
\centerline{\epsfig{file=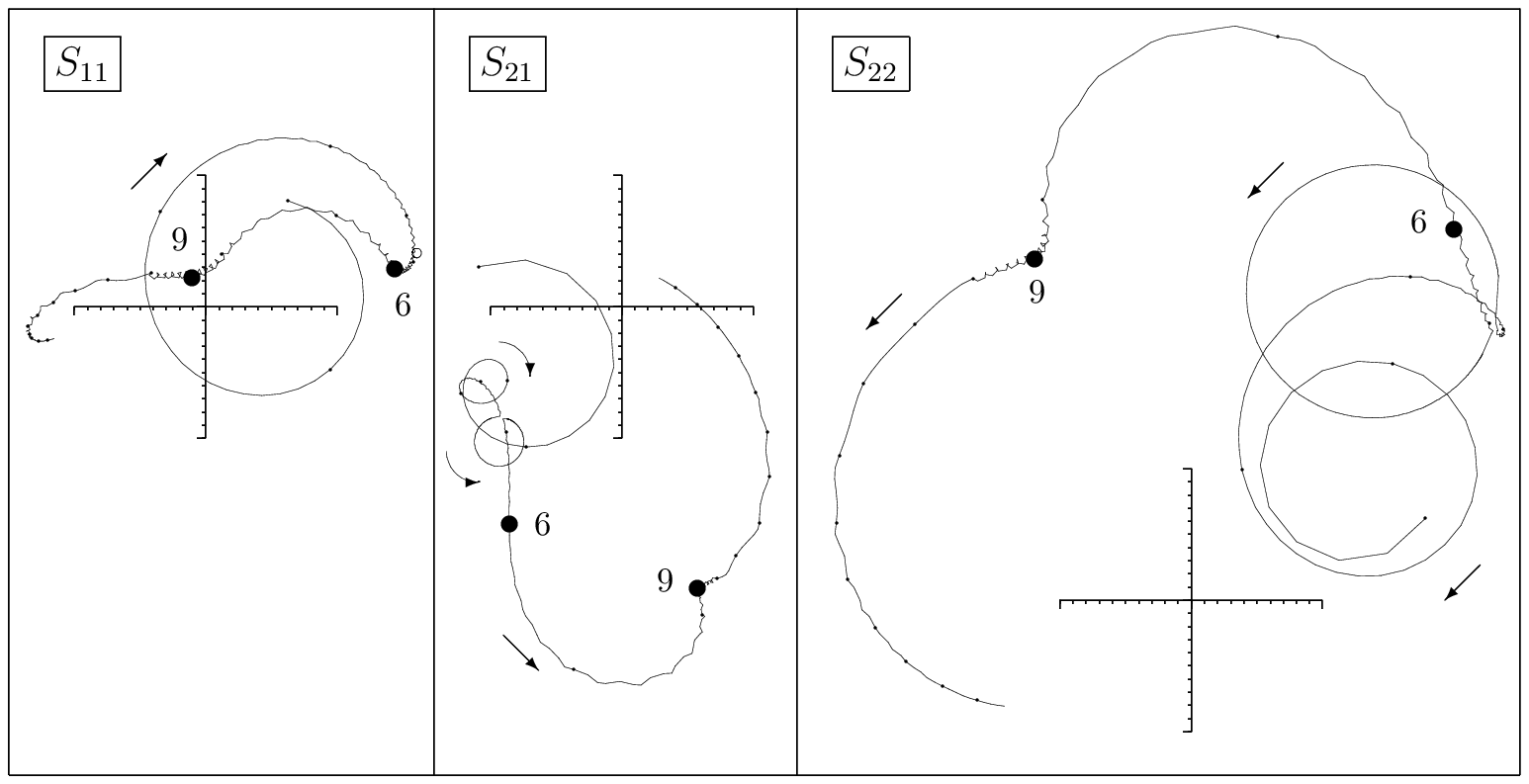}}
\caption{\sf
Argand plot of the $S$-matrix with the third pole excluded. 
The dots mark the points corresponding to  $E=6$ and $E=9$.
}
\label{fig.argand.3out}
\end{figure}

In order to examine this, we restore the second pole and exclude the
third resonance pole from the expansion (\ref{MLexpansion}). The
resulting Argand plots are shown in Fig. \ref{fig.argand.3out}. As was
expected, 
the second anti-clockwise arc 
($6<E<9$) came back together with the second pole. 
The third arc 
($E>9$) however did not disappear together with the third
pole. Therefore the third pole is not fully responsible for this arc
and thus for the wide bump on the cross sections beyond $E\sim 9$.

This bump, which is seen in all three Figs. \ref{fig.sigma11},
\ref{fig.sigma12}, and \ref{fig.sigma22}, is a collective effect of
several resonances, namely, the third, fourth, and so on.  With the
presence of the third pole, the anti-clockwise trajectories for $E>9$
on the Argand plots of $S_{21}(E)$ and $S_{22}(E)$ form closed
loops. When this pole is excluded, the curvature of these loops is
reduced to the extent that the closed loops become just open
arcs. Since they do not disappear entirely, we conclude that the
fourth and the subsequent poles also contribute to their formation,
although not so much as the third pole does.

\begin{figure}[htp]
\centerline{\epsfig{file=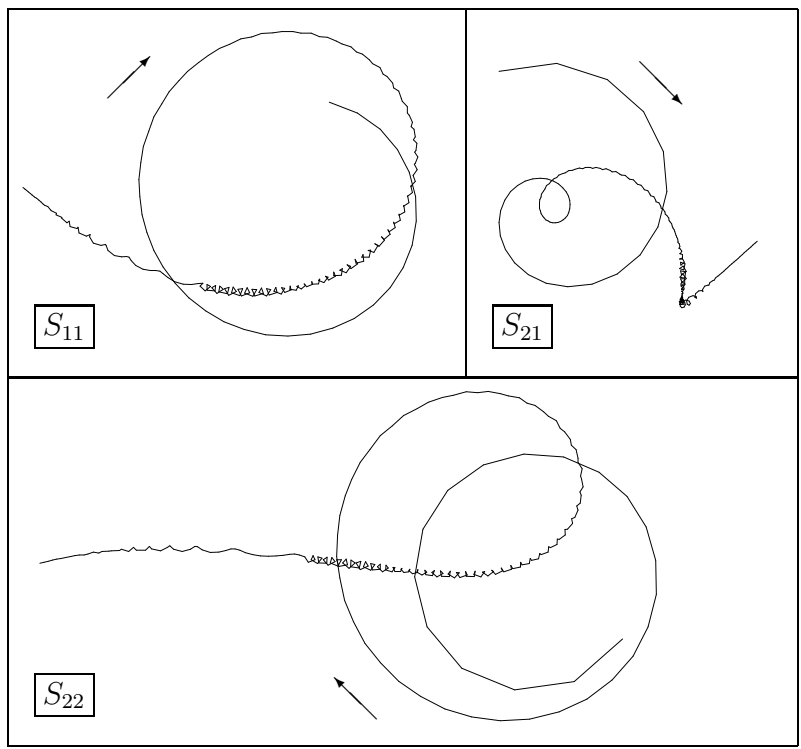}}
\caption{\sf
Argand plot of the $S$-matrix with all nine resonance poles excluded. 
}
\label{fig.argand.all_out}
\end{figure}

In order to convince ourselves that the two bumps on the cross
sections between $E\sim 6$ and $E\sim 20$ are of pure resonant nature,
we excluded all nine poles from the sum (\ref{MLexpansion}), i.e. left
only the background integral term. The Argand plots thus obtained are
given in Fig. \ref{fig.argand.all_out}.  It is seen that they show no
anti-clockwise motion.  Only close to $E=20$ the Argand trajectories
become flat and perhaps change sign of the curvature.  This means that
the integral term practically does not contribute to formation of the
bumps. It takes into account all the other poles that are outside the
integration contour. The Argand plots given in
Fig. \ref{fig.argand.all_out}, show that these outside poles
contribute very lightly at higher energies. The flattening of the
curves is the indication of this.

Finally, it is worthwhile to say a few words about the requirement of
unitarity which is violated when one analyzes the Argand plots using a
truncated Mittag-Leffler expansion.  Since the physical $S$-matrix is
unitary, its matrix elements cannot be outside the unitary circle
(circle of unit radius) on the Argand plots. This is indeed the case
with the curves given in Figs.  \ref{fig.argand.11},
\ref{fig.argand.21}, and \ref{fig.argand.22}. However when we omit
some terms in the expansion (\ref{MLexpansion}), the unitarity is
violated. As a result the Argand trajectories are not bound to
the unitary circle anymore (see Figs. \ref{fig.argand.2out} and
\ref{fig.argand.3out}). This should not be an obstacle. Actually, we
are not interested in absolute position of the curve in the complex
plane. What we look for are the segments of anti-clockwise motion. In
fact, when in our example all nine resonance poles are excluded the
unitarity is violated so much that the Argand trajectories shift very
far from the origin of the complex plane. This is why in
Fig. \ref{fig.argand.all_out} we do not indicate the coordinate system
(which in this case does not give any meaningful information anyway).

\section{Conclusion}
The method presented in this paper, for solving a two-channel problem
is based on first order differential equations that are equivalent to
the corresponding Schr\"odinger equation but are more convenient for
dealing with the resonant phenomena. The main advantage of this method
is that it enables us to directly calculate the Jost matrix for
practically any complex value of the energy. The spectral points
(bound and resonant states) can therefore be located in a rigorous
way  as zeros of the Jost matrix determinant.  

When calculating the Jost matrix, we solve differential equations and
thus at the same time obtain the wave function with the correct
asymptotic behaviour that is embedded in the solution from the outset
in an analytical form. The method therefore gives a complete solution
of the problem in one run.  Another advantage of the method used in
this paper, is that it offers very accurate way of calculating not
only total widths of resonances but their partial widths as well.

Although in this paper we deal with a two-channel problem, all the
formulae can be easily generalized to any number of channels. For
this, we only need to change the matrix dimensions from 2 to an
appropriate number $N$. 

Charged particles can also be considered within this method with minor
modifications. For this, we need to use the Coulomb functions
$F_{\ell}(\eta,kr)$ and $G_{\ell}(\eta,kr)$ instead of the
Rikkati-Bessel and Riccati-Neumann functions $j_\ell(kr)$ and
$n_\ell(kr)$, and their combinations $F_{\ell}(\eta,kr)\mp
iG_{\ell}(\eta,kr)$ instead of the Riccati-Hankel functions
$h_\ell^{(\pm)}(kr)$ when constructing the wave functions and deriving
the equations. For $\eta\to0$ such equations will be automatically
transformed into the equations given here.

The possibility of calculating the Jost matrix and the $S$-matrix for
any complex energy in the resonance domain, enables us not only to
locate the $S$-matrix resonance poles but to calculate its residues at
these poles. This, in turn, makes it possible to obtain the
Mittag-Leffler type expansion of the $S$-matrix as a sum of the
singular terms (representing the resonances) and the background term
(contour integral). Such an expansion is very useful in analyzing the
contribution of individual resonances into the scattering.

The analysis of the cross section turned out to be less informative
than expected.  The reason for this is that the cross section being a
quadratic function of the $S$-matrix, involves all possible
interference terms which smear the contributions from individual
resonances and make it difficult to clearly separate them.  However,
using the Argand plot technique, one can, as demonstrated above,
elucidate the origin of the structures in a computed cross section as
being the effects of a single isolated resonance, a collective
contribution of a set of resonances, or a non-resonant background
scattering (which in fact is a collective contribution from all far
poles of the $S$-matrix).

The two-channel model by Noro and Taylor, considered in this paper as
an example, exhibits these difficulties in an extreme form. This is
because its spectrum involves a string of almost completely
overlapping resonances. Even in such an extreme case, with the
proposed approach, we managed to assign the irregularities of the
cross section to the resonances and the background term.

Although this model might look artificial, the spectrum of resonances
generated by the Noro and Taylor potential, is similar to the spectra
of some real atomic systems. For example, the existence of a string of
overlapping resonances that starts with a narrow one, is a feature
that characterizes the charge-transfer reaction $N^{3+} + H \to N^{2+}
+ H^{+}$ \cite{R:McCarrollVa80,R:BaranyBRrand83,R:RittbyEla84}. The
analysis of the Noro and Taylor model, given in the present paper,
could therefore shed some light on the problems associated with
realistic systems.  Furthermore, with small numerical modifications,
the methods presented here can be directly applied to this
charge-transfer reaction.  The contribution from each resonant
eigenstate can then be quantified by using the corresponding $S$-matrix
residue and the Argand plots as demonstrated in this paper.

Such an analysis, may eventually, contribute to the understanding of
the $FHD$ resonances
\cite{R:Schatz2000,R:SkodjeSkouJCP2000,R:FardaySoc04} discussed in the
Introduction. Finally, the complex interference between different
resonant states in building the observable quantity - the cross
section - tells us that it is absolutely necessary to have theoretical
tool, like the present, in order to analyze the high-resolution,
low-energy atom-atom, atom-ion and atom-molecule collision experiments
that eventually may be performed \cite{R:DESIREE}.


\section*{Acknowledgements}
Financial support from the University of South Africa, the National
Research Foundation of South Africa (NRF grant number 2067421),
as well as from the STINT (Sweden) is greatly appreciated.
One of the authors (SAR) is thankful to Stockholm University for
its warm hospitality.


%
%
\begin{table}
\begin{tabular}{|r|r|r|r|r|}
\hline
$E_r$ & $\Gamma$ & $\Gamma_1$ & $\Gamma_2$ & Ref.\\
\hline
 -2.314391 & 0 & 0 & 0 & this work\\
\hline
 -1.310208 & 0 & 0 & 0 & this work\\
\hline
 -0.537428 & 0 & 0 & 0 & this work\\
\hline
 -0.065258 & 0 & 0 & 0 & this work\\
\hline
4.768197 & 0.001420 & 0.000051 & 0.001369 & this work\\
4.768197 & 0.001420 & 0.000051 & 0.001369 & \cite{Masui}\\
4.7682   & 0.001421 & 0.000061 & 0.001360 & \cite{NoroTaylor}\\
4.7682   & 0.001421 & 0.000051 & 0.001368 & \cite{MoiseevPeskin}\\
\hline
7.241200 & 1.511912 & 0.363508 & 1.148404 & this work\\
7.241200 & 1.511912 & 0.363507 & 1.148405 & \cite{Masui}\\
\hline
8.171217 & 6.508332 & 1.596520 & 4.911812 & this work\\
8.171216 & 6.508332 & 1.596517 & 4.911814 & \cite{Masui}\\
\hline
8.440526 & 12.562984 & 3.186169 & 9.376816 & this work\\
8.440526 & 12.56299 & 3.186167 & 9.376820 & \cite{Masui}\\    
\hline
8.072643 & 19.145630 & 4.977663 & 14.167967 & this work\\
8.072642 & 19.14563 & 4.977660 & 14.16797 & \cite{Masui}\\    
\hline
7.123813 & 26.025337 & 6.874350 & 19.150988 & this work\\
7.123813 & 26.02534 & 6.874348 & 19.15099 & \cite{Masui}\\ 
\hline
5.641023 & 33.070140 & 8.816746 & 24.253394 & this work\\
5.641023 & 33.07014 & 8.816744 & 24.25340 & \cite{Masui}\\     
\hline
3.662702 & 40.194674 & 10.768894 & 29.425779 & this work\\
3.662702 & 40.19467 & 10.76889 & 29.42578 & \cite{Masui}\\
\hline
1.220763 & 47.339350 & 12.709379 & 34.629971 & this work\\
1.220763 & 47.33935 & 12.70925 & 34.63010 & \cite{Masui}\\
\hline
-1.657821 & 54.460303 & 14.624797 & 39.835506 & this work\\
-1.658115 & 54.46087 & 14.62500 & 39.83587 & \cite{Masui}\\
\hline
-4.949904 & 61.523937 & 16.507476 & 45.016461 & this work\\
-4.950418 & 61.52509 & 16.50735 & 45.01774 & \cite{Masui}\\    
\hline
-8.635366 & 68.503722 & 18.352084 & 50.151638 & this work\\
-8.635939 & 68.50621 & 18.35089 & 50.15532 & \cite{Masui}\\    
\hline 
-12.696283 & 75.378773 & 20.155213  & 55.223560 & this work\\    
\hline
-17.117760 & 82.129712 & 21.915313 & 60.214399  & this work\\    
\hline
\end{tabular}
\caption{
Spectral points generated by the potential (\ref{NTpotential}) and
shown in  Fig.~\protect\ref{fig.spectrum}.   
}
\label{table.spectrum}
\end{table}
\begin{table}
\begin{tabular}{|c|r|}
\hline
    & ${\rm Res}\left[S_{11}(E)\right]$ \\
$E$ & ${\rm Res}\left[S_{21}(E)\right]$ \\
    & ${\rm Res}\left[S_{22}(E)\right]$ \\ 
\hline
                     &$(-0.04757+i0.50890)\times10^{-4}$ \\
$4.768197-i0.000710$ & $(0.26429-i0.02943)\times10^{-3}$  \\
                     & $(-0.04234-i0.13016)\times10^{-2}$ \\
\hline
                     & $-0.52405-i0.05745$ \\
$7.241200-i0.755956$ & $0.90478+i0.25589$  \\
                     & $-1.50601-i0.71154$ \\
\hline
                     & $4.17442-i0.17731$  \\
$8.171217-i3.254166$ & $-7.30357-i0.80750$ \\
                     & $12.42094+i3.31244$  \\
\hline
                     & $-4.97737+i10.97961$ \\
$8.440526-i6.281492$ & $10.99059-i17.56517$ \\
                     & $-22.57772+i27.36750$ \\
\hline
                     & $-19.17840-i7.93820$ \\
$8.072643-i9.572815$ & $30.61007+i17.10104$ \\
                     & $-47.96303-i34.49489$ \\
\hline
                      & $0.53670-i26.81386$ \\
$7.123813-i13.012669$ & $-5.46659+i44.46505$\\
                      & $16.42995-i72.88532$ \\
\hline
                      & $24.25310-i16.59884$ \\
$5.641023-i16.535070$ & $-42.53890+i23.84558$ \\
                      & $73.67460-i33.28442$ \\
\hline
                      & $28.88187+i4.77348$ \\
$3.662702-i20.097337$ & $-46.99352-i11.58529$\\
                      & $75.99725+i24.95204$ \\
\hline
                      & $19.53089+i19.50318$ \\
$1.220763-i23.669675$ & $-29.97320-i34.31696$\\
                      & $45.62484+i59.78651$ \\
\hline
\end{tabular}
\caption{
Residues of the $S$-matrix elements at the 9 resonance poles enclosed
in the contour shown in Fig.~\protect\ref{fig.contour}.   
}
\label{table.residues}
\end{table}
%
%
\newpage
\newpage
\end{document}